\documentclass[twocolumn,reprint,aps,floatfix,dvipdfmx]{revtex4-1}
\usepackage{amsmath,amssymb}
\usepackage{graphicx}
\usepackage{bm}
\usepackage{braket}
\usepackage{physics}
\usepackage{color}
\usepackage{dcolumn}
\usepackage{ulem}
\usepackage[
colorlinks=true, citecolor=blue, urlcolor=blue, linkcolor=blue,
setpagesize=false, bookmarks=false
]{hyperref}

\begin{document}

\title{
  Pump-probe spectroscopy of the one-dimensional extended Hubbard model\\at half filling
}

\author{Koudai Sugimoto$^{1,}$}
\email{sugimoto@rk.phys.keio.ac.jp}
\author{Satoshi Ejima$^{2}$}
\affiliation{
$^1$Department of Physics, Keio University, Yokohama 223-8522, Japan\\
$^2$Institut f\"{u}r Softwaretechnologie, Abteilung High-Performance Computing, Deutsches Zentrum f\"{u}r Luft- und Raumfahrt (DLR), 22529 Hamburg, Germany
}

\date{\today}

\begin{abstract}
By utilizing time-dependent tensor-network algorithms in the infinite matrix-product-state representation, we theoretically investigate the pump-probe spectroscopy of the one-dimensional extended Hubbard model at half filling.
Our focus lies on nonequilibrium optical conductivity and single-particle excitation spectra.
In the spin-density-wave (SDW) phase, we identify an in-gap state in the nonequilibrium optical conductivity due to the formation of excitons (or doublon-holon pairs), generated by the pulse through nonlocal interactions.
In the strong-coupling regime, we discern additional multiple in-gap and out-of-gap states.
In the charge-density-wave (CDW) phase, we detect not only an in-gap state but also a finite Drude weight, which results from the dissolution of charge order by photoexcitation.
Analyzing time-dependent single-particle excitation spectra directly in the thermodynamic limit confirms the origin of these new states in the SDW and CDW phases as the excitation of newly emerged dispersions.
We illustrate that the pump-probe spectroscopy simulations in the thermodynamic limit furnish unambiguous spectral structures that allow for direct comparison with experimental results, and the integration of nonequilibrium optical conductivity and time- and angle-resolved photoemission spectroscopy provides comprehensive insights into nonequilibrium states.
\end{abstract}

\maketitle

\section{Introduction}

Recent years have seen significant progress in studying novel phenomena in materials driven out of equilibrium by high-intensity laser irradiation \cite{Torre2021RMP, Dong2022AM}.
Emergent phenomena, such as photoinduced (photoenhanced) superconductivity \cite{Kaiser2017PS,Cavalleri2018CP,Fausti2011S,Suzuki2019CP,Isoyama2021CP,Buzzi2020PRX,Werner2018PRB,Wang2018PRL,Bittner2019JPSJ,Schlawin2019PRL,Werner2019PRB} and optical control of magnetic order \cite{Kirilyuk2010RMP,Ishihara2019JPSJ,Mikhaylovskiy2015NC,Baierl2016NP,Afanasiev2019PRX,Schlauderer2019N,Mentink2015NC,Li2018NC} in strongly correlated electron systems, have also started to attract both experimental and theoretical attention.
In this context, pump-probe spectroscopy has been a crucial tool for exploring nonequilibrium phenomena in these materials.
This technique involves exciting a system with a high-intensity pump pulse, followed by examining the dynamical properties of the nonequilibrium state through the linear response of a low-intensity probe pulse.

Analyzing the nonequilibrium state of strongly correlated systems within a one-dimensional (1D) model provides a beneficial starting point since the theoretical treatment of a 1D system is simpler compared to two- or three-dimensional systems.
Despite being a special case, essential characteristics of correlated systems can be elucidated from 1D systems.
Furthermore, the 1D Mott insulator, which manifests in actual materials and has been thoroughly studied in the literature, is of particular interest.
Ultrafast phenomena of 1D Mott insulators, such as organic salt ET-F$_2$TCNQ \cite{Okamoto2007PRL, Uemura2008JPSJ, Wall2011NP, Miyamoto2019CP,Takamura2023PRB} and halogen-bridged transition-metal compounds \cite{Iwai2003PRL, Matsuzaki2006JPSJ, Matsuzaki2014PRL}, have been intensively explored in preceding experiments because their optical response is significantly influenced by the formation of doublon-holon bound states due to nonlocal interactions resulting from photoexcitation.

From a theoretical perspective, photoinduced nonequilibrium states in the model with nonlocal interaction have been discussed in the literature \cite{Gomi2005PRB, Al-Hassanieh2008PRL, Takahashi2008PRB, Lu2012PRL, Bittner2019JPSJ, Shao2019PRB, Murakami2022CP, Murakami2023PRL}. 
Numerical simulations of pump-probe spectroscopy have reported the emergence of light-induced in-gap states in the half-filled 1D extended Hubbard model (1DEHM) including nearest-neighbor Coulomb interaction.
This was achieved by examining the nonequilibrium optical conductivity using time-dependent exact diagonalization \cite{Lu2015PRB} and density-matrix renormalization group \cite{Rincon2021PRB} techniques.
However, previous studies have shown system-size dependencies in their results since their calculations were conducted using finite clusters.

In this paper, by employing time-dependent tensor-network algorithms in the infinite matrix-product state (iMPS) representation, we simulate the nonequilibrium dynamics of the 1DEHM directly in the thermodynamic limit to investigate nonequilibrium phenomena induced by an intense, short-time pulse.
We elucidate the dynamical properties of this system by examining the linear response of a subsequent weak probe pulse, corresponding to pump-probe spectroscopy experiments.
Our focus lies primarily on optical conductivity and time- and angle-resolved photoemission spectroscopy (TARPES) in nonequilibrium situations.

Calculating with an infinite system enables distinct differentiation between excitation spectra arising from continuous and discrete levels.
In finite-system analyses, spectra may appear discretized even if the excitation derives from continuous levels.
By directly simulating an infinite system, this issue can be circumvented.
Additionally, thanks to the high-resolution calculation concerning momentum, we can easily obtain detailed peak structure and dispersion relations, facilitating unambiguous comparisons between theory and experiments.
Furthermore, we demonstrate that the nonequilibrium dynamics can be comprehensively understood by studying both optical conductivity and TARPES in a complementary fashion. 
Our study aims to deepen the understanding of nonequilibrium phenomena in strongly correlated electron systems and provide valuable insights for experimental investigations.

The remainder of this paper is organized as follows.
In Sec.~\ref{Sec:dK1xfUty}, we introduce the 1DEHM and describe the numerical method used in this paper.
In Sec.~\ref{sec:N0PDgf2Q}, we present the numerical results of nonequilibrium optical conductivity induced by an intense pump pulse.
In Sec.~\ref{sec:QzndfTzv}, we demonstrate the simulated single-particle excitation spectra, which are expected to be observed in TARPES experiments.
Finally, we provide conclusions and future outlook in Sec.~\ref{sec:2XWRnsey}.

\section{Model and method}
\label{Sec:dK1xfUty}

In this section, we introduce the Hamiltonian of the 1DEHM and provide a brief explanation of the numerical calculations that we performed.

\subsection{Extended Hubbard model}

We consider the 1DEHM at half filling.
Under the influence of a spatially uniform vector potential $A(t)$, the Hamiltonian is written as
\begin{align}
  \hat{H} (t)
    &= -t_{\mathrm{h}} \sum_{j, \sigma} \qty(e^{i \frac{ae}{\hbar c} A(t)} \hat{c}^{\dagger}_{j,\sigma} \hat{c}_{j+1, \sigma} + \mathrm{H.c.})
\notag \\
    &\quad + U \sum_{j} \qty(\hat{n}_{j,\uparrow} - \frac{1}{2}) \qty(\hat{n}_{j,\downarrow} - \frac{1}{2})
\notag \\
    &\quad + V \sum_{j} \qty(\hat{n}_{j} - 1) \qty(\hat{n}_{j+1} - 1),
\label{eq:Hamiltonian_of_1DEHM}
\end{align}
where $\hat{c}_{j,\sigma}$ ($\hat{c}^\dagger_{j,\sigma}$) is the annihilation (creation) operator of an electron at site $j$ with spin $\sigma$, $t_{\mathrm{h}}$ is the hopping integral, $U$ is the on-site interaction, and $V$ is the intersite interaction.
We define the number operators of the electrons as $\hat{n}_{j,\sigma} = \hat{c}^{\dagger}_{j,\sigma} \hat{c}_{j,\sigma}$ and $\hat{n}_{j} = \sum_{\sigma} \hat{n}_{j,\sigma}$.
For the sake of simplicity, the lattice constant $a$, the electron charge $-e$, the Planck constant $\hbar$, and the speed of light $c$ are set at unity, hereafter.
In the strong-coupling limit, where $U, V \gg t_{\mathrm{h}}$, the ground state (GS) of this model exhibits a spin-density wave (SDW) state for $U \gtrsim 2V$ and a charge-density wave (CDW) state for $U \lesssim 2V$ \cite{Ejima2007PRL}.
Note that the SDW-CDW transition occurs at $V / t_{\mathrm{h}} \approx 5.124$ for $U / t_{\mathrm{h}} = 10$.
In the following, we set $t_{\mathrm{h}} = 1$ as the energy unit.

\subsection{Time evolution}

We simulate the time-dependent quantum state under a high-intensity laser pulse with a vector potential given by
\begin{equation}
  A(t) = A_{0} e^{-(t - t_{0})^2/2\sigma_{0}^2} \cos(\omega_{0} t),
\end{equation}
where $A_0$ is the amplitude, $t_0$ is the central time, $\sigma_0$ is the width, and $\omega_0$ is the frequency of the pump light.
To numerically calculate the GS and the time-evolution dynamics, we employ the infinite time-evolving block decimation (iTEBD) method \cite{Vidal2007PRL,Orus2008PRB}.
The quantum state at time $t$ is denoted as $\ket{\psi (t)}$, and we set $\ket{\psi(-\infty)}$ as the GS obtained by carrying out the imaginary-time evolution.
We represent the time-evolution operator from $t'$ to $t$ as $\hat{U} (t,t') = \mathcal{T} \exp(-i \int^{t}_{t'} \dd{t''} \hat{H} (t''))$, where $\mathcal{T}$ is the time-ordering operator, allowing us to write $\ket{\psi(t)} = \hat{U} (t,-\infty) \ket{\psi(-\infty)}$.
We denote the expectation value at time $t$ as $\ev{\cdots}_{t} = \ev{\cdots}{\psi(t)}$.

In the following section, we investigate nonequilibrium dynamics when a pump pulse is applied to both the SDW and CDW states of the 1DEHM.
Namely, we focus on nonequilibrium optical conductivity and single-particle excitation spectra.
For pump-pulse parameters, we fix $\sigma_0 = 0.5$ and set $A_{0}=0.3$ for $V\neq0$ and $A_{0}=0.6$ for $V=0$.
To efficiently generate nonequilibrium states, $\omega_{0}$ is set to the value where the optical conductivity in the GS becomes the largest for each $V$.
Optical conductivities in the GS for various $V$ are given in the next section.

In the time-evolution calculations for optical conductivity and single-particle excitation spectra, we set the time step to $\delta t = 0.01$ and $0.05$, respectively, and the bond dimensions to $\chi = 3000$ and $1500$, respectively. We apply the fourth-order Trotter decomposition for optical conductivity and the second-order one for single-particle excitation spectra.

\section{Nonequilibrium optical conductivity}
\label{sec:N0PDgf2Q}

We estimate the optical conductivity by examining the response of an electric current to a weak electric field.
The current operator in a vector potential $A(t)$ is written as
\begin{align}
  \hat{J}_{A(t)}
    = - \pdv{\hat{H}}{A}
    = t_{\mathrm{h}} \sum_{j, \sigma} \qty(i e^{i A(t)} \hat{c}^{\dagger}_{j,\sigma} \hat{c}_{j+1, \sigma} + \mathrm{H.c.}).
\end{align}
Upon applying a weak electric field $E_{\mathrm{pr}}(t) = -\pdv{A_{\mathrm{pr}}(t)}{t}$ as a probe pulse in addition to the pump pulse, the induced deviation in the current per site satisfies
\begin{equation}
  j_{\mathrm{pr}} (t)
    = \frac{1}{L} \qty(\langle \hat{J}_{A + A_{\mathrm{pr}}} \rangle_t - \langle \hat{J}_{A} \rangle_t)
    = \int^{t}_{-\infty} \sigma(t,t') E_{\mathrm{pr}} (t') \dd{t'},
    \label{eq:jpr}
\end{equation}
where $L$ is the system size and $\sigma(t,t')$ represents the linear-response function for the electric field.
Since the response function should satisfy causality, i.e., $\sigma(t,t') = \theta(t-t') \sigma(t,t')$, Eq.~\eqref{eq:jpr} can be expressed as $j_{\mathrm{pr}} (t) = \int^{\infty}_{-\infty} \sigma(t,t') E_{\mathrm{pr}} (t') \dd{t'}$.
Taking the Fourier transform of both sides with respect to $t$, we obtain $j_{\mathrm{pr}} (\omega) = \int^{\infty}_{-\infty} \sigma (\omega,t') E_{\mathrm{pr}} (t') e^{i\omega t'} \dd{t'}$, where $\sigma (\omega,t') \equiv \int^{\infty}_{-\infty} \sigma (t,t') e^{i\omega (t-t')} \dd{t}$.
Assuming that the probe pulse $E_{\mathrm{pr}} (t)$ is nonzero only within the period $t_{\mathrm{pr}} \pm \tau / 2$, where $\tau$ is much smaller than the characteristic time scale of the system, and that $\sigma (\omega,t')$ remains constant during this period, we approximate $j_{\mathrm{pr}} (\omega) \simeq \sigma(\omega,t_{\mathrm{pr}}) \int^{\infty}_{-\infty} E_{\mathrm{pr}} (t') e^{i\omega t'} \dd{t'} = \sigma(\omega,t_{\mathrm{pr}}) E_{\mathrm{pr}} (\omega)$ \cite{Shao2016PRB}.
In this case, the optical conductivity of the frequency $\omega$ at probe time $t_{\mathrm{pr}}$ can be evaluated from
\begin{align}
  \sigma(\omega,t_{\mathrm{pr}})
    = \frac{j_{\mathrm{pr}} (\omega)}{i(\omega+i\eta) A_{\mathrm{pr}} (\omega)},
\label{eq:nop}
\end{align}
where $A_{\mathrm{pr}}(\omega) = E_{\mathrm{pr}}(\omega) / i(\omega + i\eta)$ with a damping factor $\eta$.
Even though this factor is introduced for the convergence of our numerical Fourier transformation, it is associated with the lifetime of the quasiparticles due to, for example, impurity scattering in actual materials.
In this way, we can directly determine equilibrium and nonequilibrium optical conductivities in the thermodynamic limit by means of the iTEBD method.
We adopt a weak, narrow probe pulse
$A_{\mathrm{pr}} (t)
  = A_{0,\mathrm{pr}}
    e^{-(t-t_{\mathrm{pr}})^2/2\sigma_{\rm pr}^2}$,
where $\omega A_{0,\mathrm{pr}}\sigma_{\mathrm{pr}} \ll 1$ is satisfied.

The advantage of this method is that it allows simultaneous calculation of the response to an external field across all frequencies.
This is because the probe pulse can be regarded as a delta function when $\sigma_{\mathrm{pr}}$ is very small compared to the characteristic time scale of the system. 
In other words, this probe pulse is a superposition of waves at all frequencies.
In particular, when assuming $\sigma_{\mathrm{pr}} \to 0$, Eq.~\eqref{eq:nop} yields the same result as the optical conductivity obtained by applying the Kubo formula, originally formulated for thermal systems, to a nonequilibrium state \cite{Shao2016PRB,Rincon2021PRB}.
Note that the same method was used to simulate the nonequilibrium optical conductivity in previous studies \cite{Lu2015PRB,Shao2016PRB,Shinjo2018PRB,Rincon2021PRB,Shinjo2022PRR}.
The ultrashort probe pulses used in this paper are idealized, and in actual pump-probe spectroscopy experiments, the probe pulse is a wave packet with a finite width.
The interpretation of nonequilibrium optical conductivity is complicated by the uncertainty relation between energy and time.
There is an ongoing debate about the theoretical description of the optical conductivity observed in actual pump-probe spectroscopy \cite{Shao2016PRB, Eckstein2008PRB, Kennes2017PRB}.
A detailed quantitative analysis to reconcile the experimental results remains a subject for future work.
In the following, we set $A_{0,\mathrm{pr}} = 0.05$, $\sigma_{\rm pr}=0.05$, and $\eta = 0.1$ for the computations of optical conductivity.
We denote $\Delta t_{\mathrm{pr}} = t_{\mathrm{pr}} - t_{0}$ and rewrite Eq.~\eqref{eq:nop} as $\sigma(\omega, \Delta t_{\mathrm{pr}})$.

\begin{figure}
  \begin{center}
  \includegraphics[width=\columnwidth]{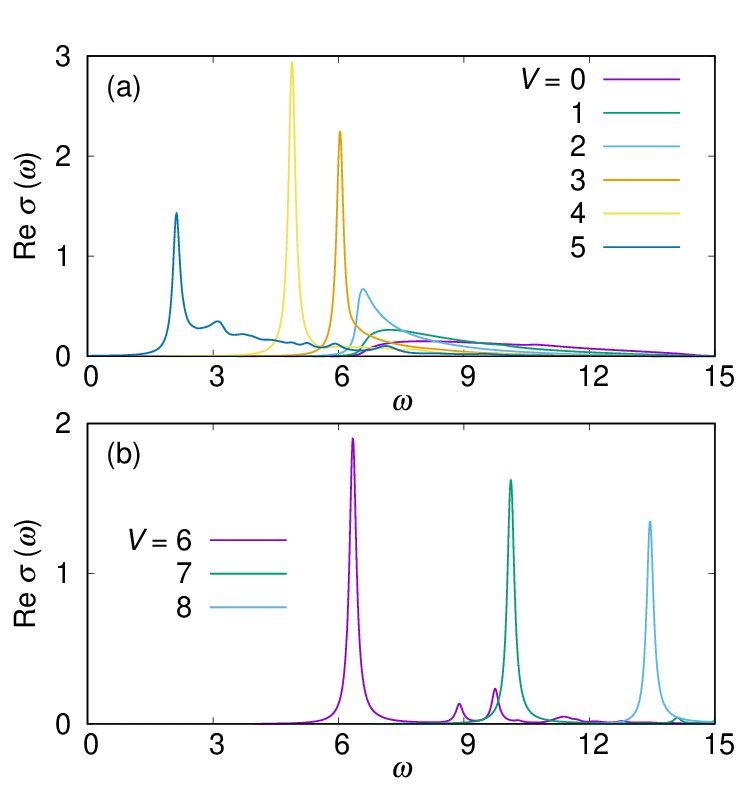}
  \end{center}
  \caption{
  The real part of the optical conductivity of the 1DEHM for $U=10$ in the GS at (a) SDW phase ($V \leq U/2$) and (b) CDW phase ($V > U/2$).
  The damping factor is set to $\eta = 0.1$.
  }\label{fig:eq-optcond}
\end{figure}

Figure~\ref{fig:eq-optcond}(a) shows the real parts of the optical conductivity of the SDW states for $U=10$ and various $V$ in the absence of the pump pulse [$A(t)=0$], which is consistent with previous dynamical density-matrix renormalization group studies \cite{Jeckelmann2002PRB, Jeckelmann2003PRB}.
A broad peak appears above a charge gap at $V=0$, originating from the excitations between the continuous levels of the upper-Hubbard band (UHB) and the lower-Hubbard band (LHB).
Turning on the intersite interaction $V$, the energy level of a doublon-holon bound state (exciton) emerges, and decreases as $V$ increases.
The energy level of the exciton becomes smaller than the bottom of the energy continuum for $V \geq 2$, leading to the emergence of a sharp peak below the charge gap \cite{Gallagher1997PRB}.
We should note that in our numerical calculations the amplitude of this excitonic peak is finite due to the finite $\eta$ and diverges as $\eta \to 0$ \cite{Jeckelmann2003PRB}.
The excitonic energy level becomes the lowest value at the SDW-CDW transition point $V \simeq U/2$.

The optical conductivities of the CDW state $V > U/2$ are shown in Fig.~\ref{fig:eq-optcond}(b).
Here, the peak position of $\Re \sigma (\omega)$ increases as $V$ increases.
This peak position corresponds to the energy required to dissociate a doublon.

The peak positions of the optical conductivities can be readily estimated in the strong-coupling limit ($U,V \gg t_{\mathrm{h}}$).
In this limit, the SDW state comprises singly occupied sites.
Thus, the energy of the first-excited state, characterized by the presence of an adjacent doublon and holon, is approximately $U-V$.
On the other hand, doublons and holons align alternately in the CDW state.
Therefore, the first-excited state, where two adjacent sites become singly occupied, requires an excitation energy of approximately $3V-U$.

\subsection{SDW phase}

\begin{figure}
  \begin{center}
  \includegraphics[width=\columnwidth]{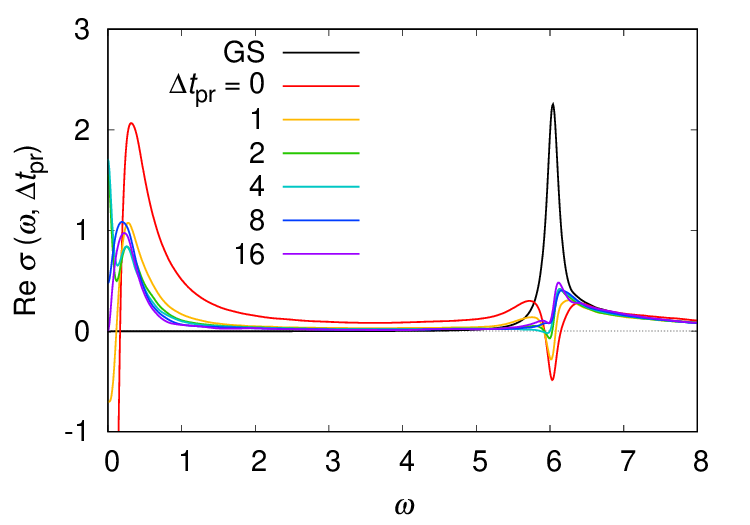}
  \end{center}
  \caption{
    The real part of nonequilibrium optical conductivity of the 1DEHM at $U=10$ and $V=3$ (SDW phase) for various probe times.
    The black solid line indicates the optical conductivity in the GS.
    The pump-light frequency and intensity are set to $\omega_0 = 6.04$ and $A_0 = 0.3$, respectively.
  }
  \label{fig:noneq-opt-cond_U10_V3}
\end{figure}

We first show the nonequilibrium optical conductivity of the 1DEHM with $U=10$ and $V=3$ in Fig.~\ref{fig:noneq-opt-cond_U10_V3}.
The pump-pulse frequency is set to $\omega_0 = 6.04$, where $\Re \sigma(\omega)$ reaches its maximum [see Fig.~\ref{fig:eq-optcond}(a)].
By comparing the spectra obtained at $\Delta t_{\mathrm{pr}} = 0$ with those in the absence of the pump pulse, we observe two characteristic features.
One is the negative spectra at the pump-light frequency $\omega = \omega_0$, which may originate from the population inversion of the electrons due to the pump-pulse irradiation.
This nonthermal state leads to stimulated emission by the probe pulse, resulting in the negative optical conductivity \cite{Ryzhii2007JAP}. 
The other is the emergence of a new peak at small $\omega$, which implies the creation of an in-gap state.
This in-gap state is associated with two types of excitons in the 1DEHM: even- and odd-parity excitons \cite{Lu2015PRB}.
The energy level of the even-parity excitons is slightly larger than that of the odd-parity excitons, and the optical transitions to the even-parity excitons from the GS are forbidden \cite{Mizuno2003PRB, Matsueda2004PRB, Yamaguchi2021PRB}.
However, once a state enters the odd-parity excitonic state due to the pump pulse, the state can be further excited to an even-parity exciton level by the subsequent probe pulse.
Therefore, transitions to the even-parity excitonic state, which are not allowed from the GS, are realized.

Once the pump pulse has passed, the system begins to relax via the recombination of doublons and holons.
As shown in Appendix \ref{app:docc}, this is evident from the gradual decrease in double occupancy.
The negative spectra at $\omega = \omega_0$ gradually turn into positive asymmetric ones, which implies the emergence of Fano resonance.
In this case, there is a quantum interference between the exciton level and the doublon-holon continuum \cite{Rincon2021PRB}.
On the other hand, the in-gap state remains over time.
The reason why the peak position with regard to the in-gap state at $\Delta t_{\mathrm{pr}}>0$ becomes smaller than that at $\Delta t_{\mathrm{pr}}=0$ can be attributed to the Stark effect of the excitons \cite{Udono2022PRB, Udono2023PRB}.
The electric field of the pump pulse at $\Delta t_{\mathrm{pr}}=0$ enhances the splitting between the even- and odd-parity exciton levels.
The energy of the in-gap state eventually stabilizes at approximately $\omega \approx 0.2$.

\begin{figure}
  \begin{center}
  \includegraphics[width=\columnwidth]{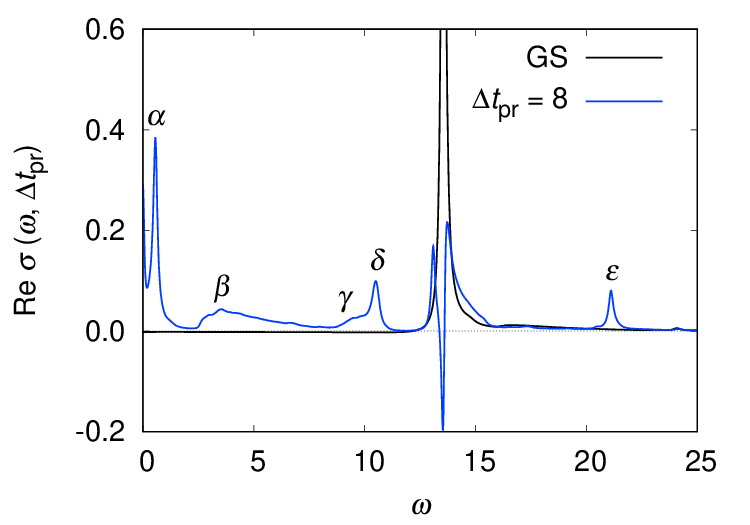}
  \end{center}
  \caption{
    The real part of the nonequilibrium optical conductivity of the 1DEHM at $U=20$ and $V=6$ (SDW phase).
    The black and blue solid lines are for the GS and the nonequilibrium state at $\Delta t_{\mathrm{pr}} = 8$, respectively.
    The labels above the spectrum denote the new peaks that arise in the nonequilibrium state.
    The pump-light frequency and intensity are set to $\omega_0 = 13.55$ and $A_0 = 0.3$, respectively.
  }
  \label{fig:noneq-opt-cond_U20_V6}
\end{figure}

To further scrutinize the aforementioned features, we also examine the SDW state with a larger interaction strength.
Figure~\ref{fig:noneq-opt-cond_U20_V6} shows the nonequilibrium optical conductivity at $U=20$ and $V=6$ with $\omega_0 = 13.55$.
In addition to the low-energy peak originating from two exciton levels at small $\omega$ \cite{Lu2015PRB}, labeled as $\alpha$, we find that there are two broad structures ($\beta$ and $\gamma$) and a sharp peak at $\omega \approx 10.5$ ($\delta$) below the optical gap.
Moreover, a new peak arises at $\omega \approx 21.1$ ($\varepsilon$), which is located at a higher energy than the exciton level.
Apart from the peak $\alpha$, the origins of these peaks can be better understood from the numerical results of nonequilibrium single-particle excitation spectra, which we will discuss in the subsequent section.

\subsection{CDW phase}

\begin{figure}
  \begin{center}
  \includegraphics[width=\columnwidth]{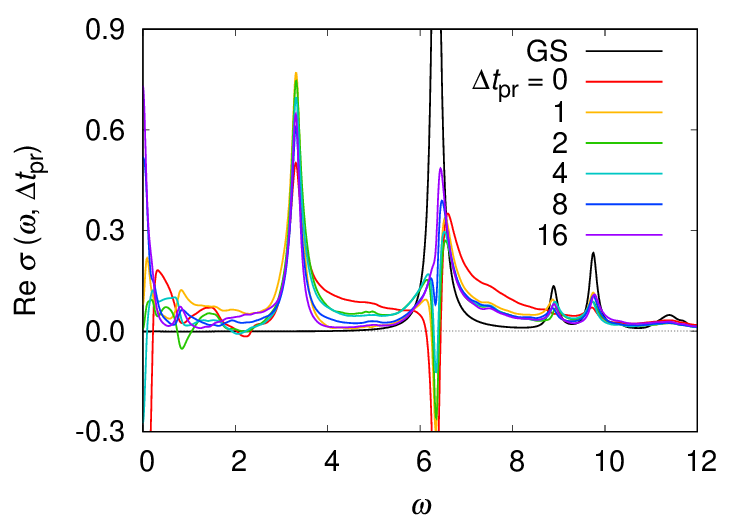}
  \end{center}
  \caption{
  The real part of the nonequilibrium optical conductivity of the 1DEHM at $U=10$ and $V=6$ (CDW phase) for various probe times.
  The black solid line indicates the optical conductivity in the GS.
  The pump-light frequency and intensity are set to $\omega_0 = 6.34$ and $A_0 = 0.3$, respectively.
  }
  \label{fig:noneq-opt-cond_U10_V6}
\end{figure}

Figure~\ref{fig:noneq-opt-cond_U10_V6} illustrates the real part of the nonequilibrium optical conductivity upon application of the pump pulse with $\omega_0 = 6.34$ to the CDW state with $U=10$ and $V=6$.
Similar to the SDW phase, the pump-pulse irradiation results in a prominent negative spectrum at $\omega = \omega_0$.
Following the passage of the pump pulse, the spectral weight at this energy recovers to positive values, eventually forming an asymmetric spectrum.

A notable change is the emergence of a new state at $\omega \approx 3.3$ in the nonequilibrium state, which is consistent with the previous study \cite{Lu2015PRB}.
This in-gap state can be interpreted from the newly formed bands in the single-particle spectra, as will be discussed later.

We also observe messy spectral structures for $\omega < 2$ that are newly generated and strongly time dependent.
At $\Delta t_{\mathrm{pr}} = 16$, the Drude weight (i.e., the spectrum at $\omega=0$) becomes finite, indicating photoinduced metallization.
Unfortunately, our iTEBD simulations are constrained to this time due to limited numerical accuracy and computational-time restriction.
Further time evolution, while maintaining accuracy, should allow the system to reach a steady state.

\section{Time- and angle-resolved photoemission spectra}
\label{sec:QzndfTzv}

We now discuss the time-dependent single-particle excitation spectra in the 1DEHM, taking into account TARPES experiments.
The intensity of single-particle excitation spectra with momentum $k$ and energy $\omega$ is given by \cite{Freericks2009PRL, Freericks2015PS}
\begin{align}
  &A^{-} (k, \omega, \Delta t_{\mathrm{pr}})
  \notag\\
  &= \frac{1}{L} \sum_{j,\ell,\sigma} e^{-ik(r_j - r_{\ell})}
      \int^{\infty}_{-\infty} \dd{t_1}
      \int^{\infty}_{-\infty} \dd{t_2}
      e^{i\omega(t_1 - t_2)}
  \notag \\
  &\quad \times
      s(t_1 - t_{\mathrm{pr}}) s(t_2 - t_{\mathrm{pr}})
      \ev{\hat{c}^{\dagger}_{\ell,\sigma} (t_2,-\infty) \hat{c}_{j,\sigma} (t_1,-\infty)}_{-\infty},
  \label{eq:TrARPES}
\end{align}
where $s(t - t_{\mathrm{pr}})$ is an envelope function of the wave packet of the probe pulse at central time $t_{\mathrm{pr}}$ and $\hat{c}_{j,\sigma} (t,t') = \hat{U}^\dagger (t,t') \hat{c}_{j,\sigma} \hat{U} (t,t')$ is the Heisenberg representation.
The envelope function utilized in this paper is a Gaussian function written as
\begin{equation}
  s(t-t_{\mathrm{pr}})
    = \frac{1}{\sqrt{2\pi} \sigma_{\mathrm{pr}}}
      \exp(-\frac{\qty(t-t_{\mathrm{pr}})^2}{2 \sigma_{\mathrm{pr}}^2}),
\end{equation}
where $\sigma_{\mathrm{pr}}$ represents the width of the wave packet.
To simulate single-particle excitation spectra in nonequilibrium, we construct the window state from the iMPS obtained from the iTEBD method and employ the infinite-boundary conditions with a uniform update scheme \cite{Zauner2015JPCM}.
For more detail, refer to Ref.~\cite{Ejima2022PRR}.
In addition, we define the integrated photoemission spectra as
\begin{equation}
  A^{-} (\omega, \Delta t_{\mathrm{pr}})
    = \int^{\pi}_{-\pi} \frac{\dd{k}}{2\pi} A^{-} (k,\omega, \Delta t_{\mathrm{pr}}),
\end{equation}
which in the following will be denoted as the time-resolved density of states (TDOS).

Increasing the width of the probe-pulse wave packet improves the energy resolution, but decreases the time resolution due to the uncertainty relation.
In this section, we set the width of the probe pulse to $\sigma_{\mathrm{pr}} = 3$.
We find that a window-state size of $L_{\mathrm{w}} = 64$ is sufficient for this case.
We present calculations for the single-particle excitation spectra of nonequilibrium states at $\Delta t_{\mathrm{pr}}=0$ and $8$.
However, we have confirmed that the spectral shape for $\Delta t_{\mathrm{pr}}>8$ is almost unchanged from that for $\Delta t_{\mathrm{pr}}=8$, implying that the single-particle excitation spectra are essentially stationary over time after photoexcitation.

\begin{figure*}
  \begin{center}
  \includegraphics[width=\linewidth]{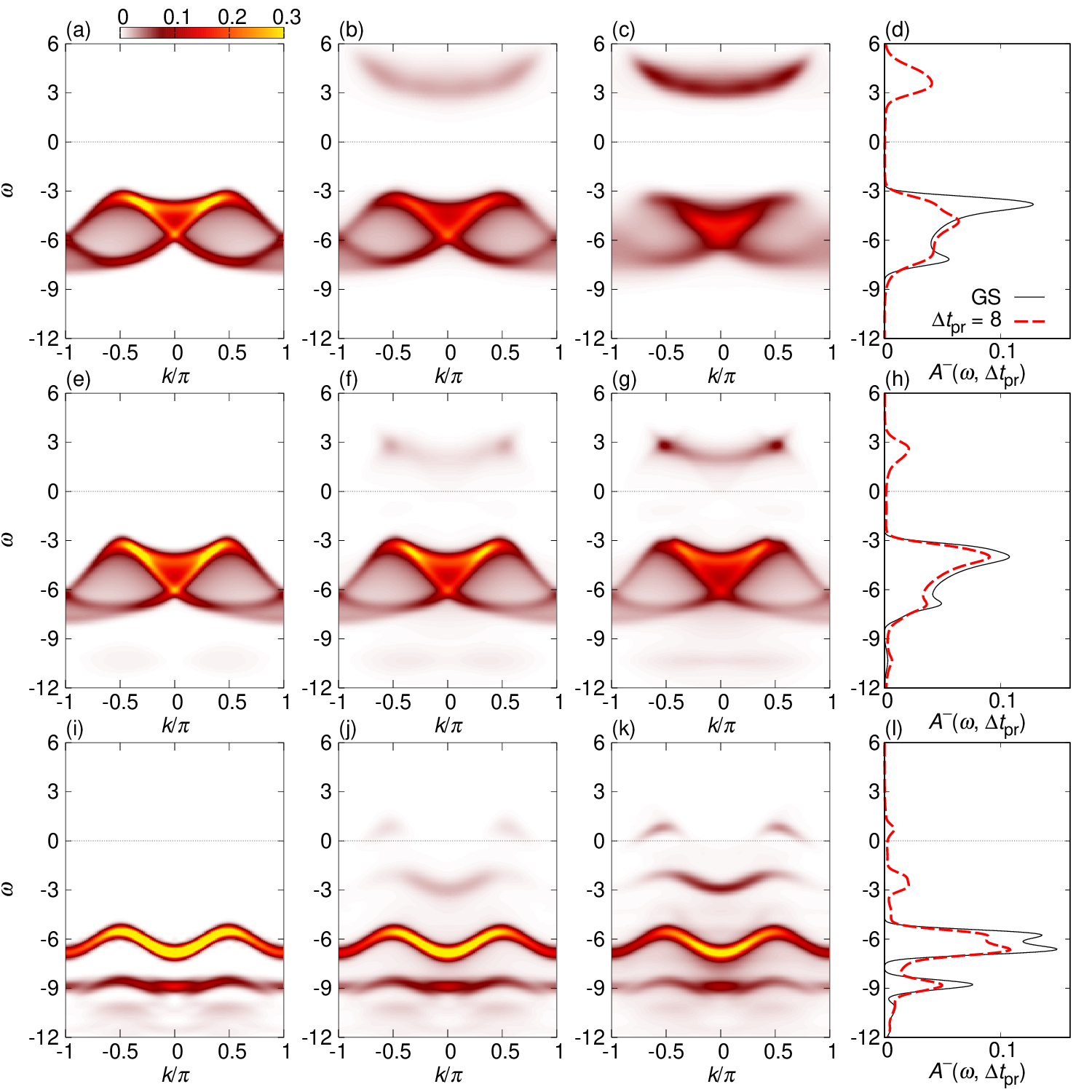}
  \end{center}
  \caption{
  Calculated single-particle excitation spectra of the 1DEHM at
    (a), (e), (i) $\Delta t_{\text{pr}} = -\infty$ (GS);
    (b), (f), (j) $\Delta t_{\text{pr}} = 0$; and
    (c), (g), (k) $\Delta t_{\text{pr}} = 8$.
  (d), (h), (l) TDOSs at $\Delta t_{\text{pr}} = -\infty$ (black solid line) and $\Delta t_{\text{pr}}=8$ (red dashed line).
  The on-site interaction is set to $U=10$, and the intersite interaction, the pump-light frequency, and its intensity are set to
    (a)-(d) $V=0$, $\omega_0 = 8.0$, and $A_0 = 0.6$;
    (e)-(h) $V=3$, $\omega_0 = 6.04$, and $A_0 = 0.3$; and
    (i)-(l) $V=6$, $\omega_0 = 6.34$, and $A_0 = 0.3$.
  }
  \label{eq:uZw2yjVp}
\end{figure*}

\subsection{SDW phase}

Let us first recall the results of single-particle excitation spectra in the pure Hubbard model, i.e., $V=0$ in Eq.~\eqref{eq:Hamiltonian_of_1DEHM}, as shown in the upper panels of Fig.~\ref{eq:uZw2yjVp}, which have also been discussed in Ref.~\cite{Ejima2022PRR}.

In the absence of the pump pulse [$A(t)=0$], the Bethe ansatz \cite{Lieb1968PRL,Essler2005} provides the exact energy dispersion, which explains the results for $U=10$ in Fig.~\ref{eq:uZw2yjVp}(a).
There are one spinon and two holon bands due to spin-charge separation.
The two holon bands are degenerate at $k=0$ and $\pm \pi$, and their width is $4 t_{\mathrm{h}}$.
The spinon and holon bands split at $k=0$, while the upper holon band merges with the spinon band at $k=\pm\pi/2$.
The spinon-holon excitation continuum below the lower holon band is visible for $\abs{k} \geq \pi/2$.
The spectra obtained here correspond to the LHB.
A detailed comparison of the exact results and the calculated spectra can be found in Refs.~\cite{Benthien2007PRB, Ejima2021SPP, Murakami2021PRB}.

We now turn to the case for the nonequilibrium state.
It should be noted that this photoexcited state is associated with the emergence of the so-called $\eta$-pairing state \cite{Kaneko2019PRL,Kaneko2020PRR,Ejima2020PRR,Ejima2020JPSCP}, which is the exact eigenstate of the Hubbard model \cite{Yang1989PRL,Essler2005}.
Figures~\ref{eq:uZw2yjVp}(b) and \ref{eq:uZw2yjVp}(c), respectively, show $A^{-} (k, \omega, \Delta t_{\mathrm{pr}})$ at $\Delta t_{\mathrm{pr}} = 0$ and $8$.
The pump pulse induces a photoexcited state, leading to the emergence of new spectral weights with a dispersion ranging from $k = -\pi$ to $\pi$ above the Fermi level and exhibiting a minimum at $k=0$.
A similar dispersion can also be observed in finite-temperature photoemission spectra attributed to thermally excited electrons \cite{Ejima2021SPP,Nocera2018PRB,Nishida2020JPSJ}.
This observation suggests that the electrons in the LHB are resonantly excited into the UHB by the pump pulse \cite{Wang2017PRB}.
Simultaneously, a reduction in the spectral intensity of the LHB occurs.
The shift of the spectral weight after pulse irradiation can also be confirmed in the results of the TDOS [see Fig.~\ref{eq:uZw2yjVp}(d)].

Next, we introduce intersite interactions.
Figure~\ref{eq:uZw2yjVp}(e) shows the single-particle excitation spectra in the GS of the 1DEHM for $U = 10$ and $V = 3$.
By introducing $V$, the charge gap becomes slightly smaller; however, the dispersion relation is almost the same as for the case of $V = 0$.
Unlike the optical-conductivity spectra, the single-particle excitation spectra in the GS do not display features associated with excitons.
This is because photoemission involves the removal of a single electron from the system, and therefore does not form a doublon-holon bound state.
We also find weak but new spectral weights around $k = \pm\pi/2$ and $\omega \approx -10.3$ appearing below the LHB.
While it is slightly difficult to see them in the intensity plot of Fig.~\ref{eq:uZw2yjVp}(e), they can be recognized in the DOS illustrated in Fig.~\ref{eq:uZw2yjVp}(h) as a black solid line.

Figures~\ref{eq:uZw2yjVp}(f) and \ref{eq:uZw2yjVp}(g) show $A^{-} (k, \omega, \Delta t_{\mathrm{pr}})$ after the pump pulse irradiation. 
In this case, the pump pulse creates numerous doublons and holons, leading to the formation of excitons due to the nonlocal interactions.
We find a new dispersion above the Fermi level, as in the case of $V = 0$.
However, unlike the case of $V = 0$, this dispersion has the maxima at $k = \pm \pi / 2$.
The difference between the maximum energy of the newly emerged band and that of the LHB is almost equal to the excitonic energy, estimated from the peak position of the optical conductivity [$\omega \approx 6.04$, see Fig.~\ref{fig:eq-optcond}(a)].
Therefore, we can interpret that the new dispersion originates from the excitons created by the pump pulse.
This new band has the same dispersion as the LHB and its visibility increases with the intensity of the pump pulse (not shown here).

\begin{figure}
  \begin{center}
  \includegraphics[width=\columnwidth]{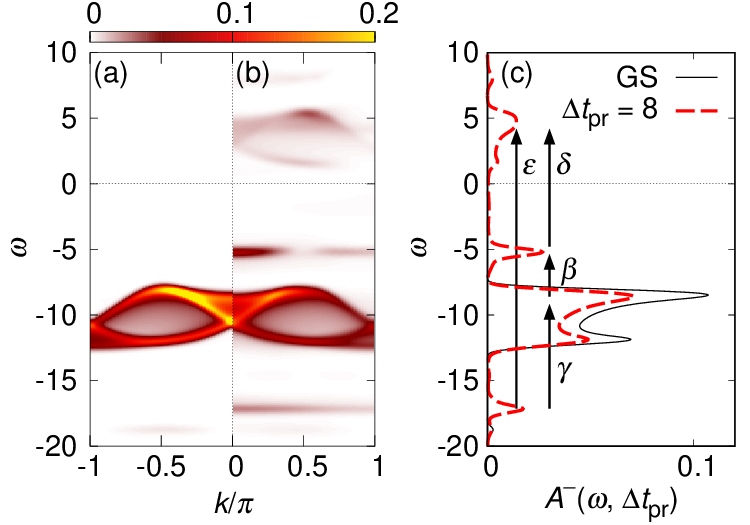}
  \end{center}
  \caption{
  Calculated single-particle excitation spectra of the 1DEHM for $U=20$ and $V=6$ at
  (a) $\Delta t_{\mathrm{pr}} = -\infty$ (GS) and
  (b) $\Delta t_{\mathrm{pr}} = 8$.
  (c) The TDOS at $\Delta t_{\mathrm{pr}} = -\infty$ (black solid line) and $\Delta t_{\mathrm{pr}} = 8$ (red dashed line).
  The arrows in the TDOS correspond to the optical excitations observed in the nonequilibrium optical conductivity in Fig.~\ref{fig:noneq-opt-cond_U20_V6}.
  The pump-light frequency and intensity are set to $\omega_0 = 13.55$ and $A_0 = 0.3$, respectively.}
  \label{fig:i0ltfr6A}
\end{figure}

To better clarify the state under the creation of excitons by the pump pulse in the SDW phase, we also present the results with larger interaction parameters.
Figure~\ref{fig:i0ltfr6A}(a) shows the single-particle excitation spectra in the GS at $U=20$ and $V=6$.
Reflecting the large interaction parameters, the LHB appears at a lower-energy level.
Figure~\ref{fig:i0ltfr6A}(b) shows $A^{-} (k, \omega, \Delta t_{\mathrm{pr}})$ after the pump-pulse irradiation, which resembles the dispersions obtained by the interaction quench \cite{Zawadzki2019PRB}.
A new dispersion with the same shape as the LHB appears above the Fermi level.
The energy difference between the new band and the LHB is the same as the excitonic energy estimated from the peak position of $\Re \sigma(\omega)$ ($\omega \approx 13.55$, see Fig.~\ref{fig:noneq-opt-cond_U20_V6}). 

Furthermore, dispersionless flat bands, which may be associated with charge-order fluctuations, also appear both above and below the LHB \cite{Lu2012PRL,Zawadzki2019PRB}.
Recall that, in the nonequilibrium optical conductivity after the pump-pulse irradiation, four peaks $\beta$, $\gamma$, $\delta$, and $\varepsilon$ emerge in addition to the peak $\alpha$ originating from the excitation between different parity excitons (see Fig.~\ref{fig:noneq-opt-cond_U20_V6}).
By comparing the peak positions in the optical conductivity with the energy-level differences of the peaks in the DOS, we can identify that the broad peak $\beta$ is ascribed to the excitation from the LHB to the flat band, the hump structure $\gamma$ originates from the excitation from the flat band to the LHB, and the two peaks $\gamma$ and $\varepsilon$ arise from the excitation from the flat bands to the newly emerging bands above the Fermi level.
The corresponding optical excitations are depicted by black arrows in Fig.~\ref{fig:i0ltfr6A}(c).

\subsection{CDW phase}

Finally, we examine the time-dependent single-particle excitation spectra in the CDW phase, as shown in the lower panels of Fig.~\ref{eq:uZw2yjVp} for $U=10$ and $V=6$.
Under the formation of charge ordering in the GS, the two holon bands, which are degenerate at $k=0$ and $\pm\pi$ in the SDW phase, split into a band with a cosine-type dispersion centered at $\omega \approx -6$ and a relatively flat band centered at $\omega \approx -9$.
Figures~{\ref{eq:uZw2yjVp}(j) and \ref{eq:uZw2yjVp}(k) show $A^{-} (k, \omega, \Delta t_{\mathrm{pr}})$ under the influence of the pump pulse.
Two new dispersions emerge around the Fermi level.
These results have been previously reported by exact diagonalization for small clusters \cite{Shao2020PRB}.
Thanks to the higher-resolution spectra obtained directly in the thermodynamic limit, the excitation from the band around $\omega \approx -6$ to the one around $\omega \approx -2.7$ can be significantly distinguished, which is related to the in-gap state observed in the nonequilibrium optical conductivity shown in Fig.~\ref{fig:noneq-opt-cond_U10_V6}.

\section{Conclusions and outlook}
\label{sec:2XWRnsey}

We explored the pump-probe spectroscopy of the 1DEHM at half filling in an infinite system.
In the strong-coupling regime, the GS of this model resides in the SDW phase when $U \gtrsim 2V$ and in the CDW phase when $U \lesssim 2V$.
In the SDW phase, doublons and holons, which are generated by the intense pump pulse, form bound states known as excitons through nonlocal interactions.
In the CDW phase, the charge order dissolves due to photoexcitation.
The dynamical response in the nonequilibrium state was revealed using the iTEBD method.

We detected an in-gap state at small $\omega$ in the nonequilibrium optical conductivity for the model with $U=10$ and $V=3$, which resides in the SDW phase.
This state can be interpreted as the transition between the odd-parity exciton and the even-parity exciton.
Furthermore, we investigated a stronger interaction model with $U=20$ and $V=6$.
In addition to this in-gap state originating from the different parity excitons, we discovered that additional peaks appear below and above the excitonic energy.
The origin of these additional peaks can be understood by examining the single-particle excitation spectra.
Specifically, the new dispersions appearing above and below the LHB after the pump pulse irradiation are associated with these new peaks in the optical conductivity.
We also observed that the LHB is replicated in a higher-energy region, where the energy difference is equal to the excitonic energy.

Moreover, we examined the nonequilibrium optical conductivity for the model with $U=10$ and $V=6$, which resides in the CDW phase.
In this case, we also found an in-gap state after the pump pulse irradiation.
Additionally, we discovered that the Drude weight becomes finite, suggesting the metallization of the system.
The origin of this in-gap state can be understood from the single-particle excitation spectra.

It would also be interesting to explore nonequilibrium physics close to SDW-CDW phase boundaries, although this paper focused on pump-probe spectroscopy deep in the SDW and CDW phases.
For instance, the time-resolved single-particle spectral function has been studied around these phase boundaries by means of the exact-diagonalization technique \cite{Shao2020PRB}.
Furthermore, a recent study indicates peculiar optical responses in high-harmonic generation near the phase boundary \cite{Shao2022PRL}.
The issue with performing iTEBD simulations near the quantum phase transition point is the increasing bond dimensions required.
It is thus highly desirable to improve the accuracy of numerical techniques and simultaneously reduce the computational time.

Lastly, we address the correspondence between our theoretical findings and experimental observations.
The 1D Mott insulator ET-F$_2$TCNQ is well described by the 1DEHM with interaction parameters $U=10$ and $V=3$ \cite{Yamaguchi2021PRB}.
In fact, the emergence of the in-gap state in optical conductivity after pump pulse irradiation has been observed.
If TARPES becomes feasible in this material, we expect that the single-particle excitation spectra obtained in this paper would also be observable.
Another approach to realizing our results is by employing a cold atomic system \cite{Bohrdt2018PRB}.
With an artificial gauge field mimicking the pump pulse, the observation of single-particle excitation spectra may be feasible.
It is worth emphasizing that our calculations, performed on an infinite system, allow for a direct comparison of the spectra observed in future experiments with our results.

In this paper, our focus was on nonequilibrium optical conductivity and single-particle excitation spectra.
Recently, time-resolved resonant inelastic x-ray scattering (RIXS) spectra have become observable through pump-probe spectroscopy \cite{Cao2019PTRSA, Mitrano2020CP, Gelmukhanov2021RMP}.
RIXS allows us to examine the dynamical correlations of charge and spin, including their momentum dependence, thereby enabling us to obtain more detailed information on strongly correlated materials.
We anticipate further developments in theoretical studies of pump-probe spectroscopy using tensor-network algorithms in the future.

\section*{Acknowledgments}
K.S. was supported by Japan Society for the Promotion of Science (JSPS) KAKENHI Grants No. JP20H01849, No. JP21K03439, and No. JP23K03286 and by Japan Science and Technology Agency (JST) COI-NEXT Program Grant No. JPMJPF2221.
S.E. was supported by the DLR Quantum Computing Initiative and the Federal Ministry for Economic Aﬀairs and Climate Action \cite{dlr}.
The iTEBD calculations were performed using the ITensor library \cite{Fishman2022SPC}.

\appendix

\section{Double occupancy}
\label{app:docc}

\begin{figure}
  \begin{center}
  \includegraphics[width=\columnwidth]{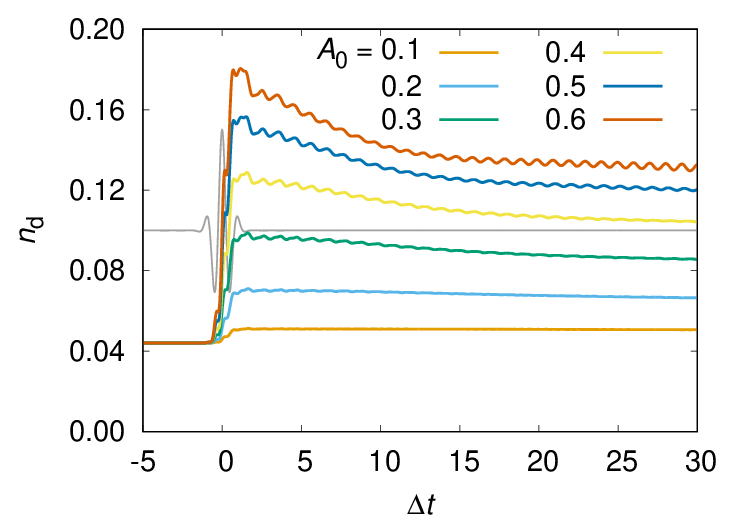}
  \end{center}
  \caption{
  The double occupancies of the 1DEHM at $U=10$ and $V=3$ for various $A_0$ as functions of time.
  The gray solid line indicates the time dependence of the pump pulse $A(t)$.
  We denote $\Delta t = t - t_0$ in the horizontal axis, where $t_0$ is the central time of the pump pulse.
  The pump-light frequency and intensity are set to $\omega_0 = 6.04$ and $A_0 = 0.3$, respectively.
  }
  \label{fig:double-occupancy}
\end{figure}

Figure \ref{fig:double-occupancy} shows the double occupancies
\begin{equation}
  n_{\text{d}} (t)
    = \frac{1}{L} \sum_j \ev{\hat{n}_{j,\uparrow} \hat{n}_{j,\downarrow}}_t
\end{equation}
of the 1DEHM at $U=10$ and $V=3$ as functions of time, for various pump-pulse intensities.
Upon the pump-pulse irradiation, doublons and holons are generated, leading to an increase in double occupancy.
Following the passing of the pump light, the system begins to relax gradually.
The relaxation process of the photoexcited 1DEHM is still unclear, but there are some suggestions, such as the Auger recombination of doublons and holons \cite{Segawa2011JPSJ,Gomi2014JPSJ}.

The larger the value of $A_0$, the more significant the increase in the double occupancy.
When a state is intensely excited by a pump pulse, the entanglement of the quantum state evolves throughout the system, suggesting that the system can no longer be described by the iMPS.
Specifically, the truncation error reaches up to $4 \times 10^{-5}$ in the calculation with $A_0=0.6$.
For $A_0=0.3$, we confirmed that the truncation error is suppressed to below $4 \times 10^{-6}$.

\section{Linear response theory of optical conductivity}

In an ideal scenario, the optical conductivity should be calculated using the Kubo formula, which is based on the linear response theory and requires the calculation of current-current correlation functions \cite{Rincon2021PRB,Shao2016PRB,Lenarcic2014PRB,Ohara2013PRB}.
This process necessitates the creation of a window state with infinite-boundary conditions \cite{Zauner2015JPCM, Ejima2022PRR} and the application of the local current operator at the center of the window state.
Given the requirement for long-time simulation to derive optical conductivities, the influence of the local current operator applied at the center site extends to the boundary before the calculation is finished.
We determined that, for a damping factor of $\eta = 0.1$, a window state of size $L_{\mathrm{w}} > 128$ should be prepared.
As this incurs substantial computational cost, we have chosen to employ the method delineated in the main text.

\section{Numerical convergence of nonequilibrium optical conductivity}

As described in the main text, we employ a fourth-order Trotter decomposition for optical-conductivity calculations.
This approach ensures the accuracy of the numerical Fourier transformation, which necessitates extended simulation time.
In this paper, the induced deviation in the current $j_{\mathrm{pr}} (t)$ is calculated up to $t - t_{\mathrm{pr}} \leq 100$.

\begin{figure}
  \begin{center}
  \includegraphics[width=\columnwidth]{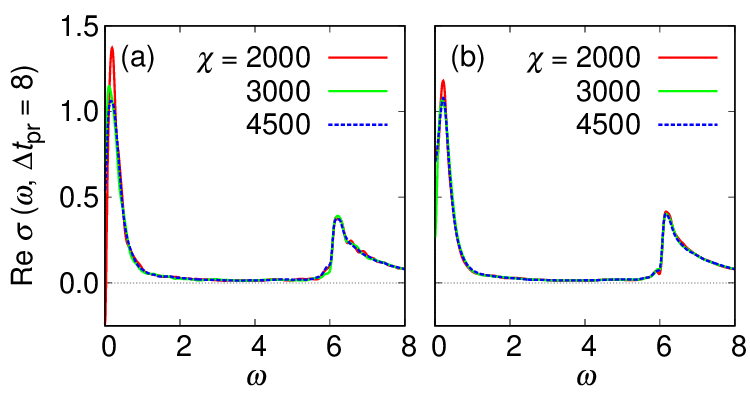}
  \end{center}
  \caption{
    The real part of the nonequilibrium optical conductivity of 1DEHM with $U=10$ and $V=3$ at $\Delta t_{\mathrm{pr}} = 8$ using (a) second-order and (b) fourth-order Trotter decompositions for various bond dimensions $\chi$.
    The pump-light frequency and intensity are set to $\omega_0 = 6.04$ and $A_0 = 0.3$, respectively.
  }
  \label{fig:trotter-and-optcond}
\end{figure}

Figure~\ref{fig:trotter-and-optcond} shows the optical conductivity calculated using both second- and fourth-order Trotter decompositions.
For optical conductivity with second-order Trotter decomposition, the calculated spectra do not converge especially at the in-gap-state energy ($\omega \approx 0.2$) even when the bond dimension is increased up to $\chi = 4500$, indicating that we do not obtain the appropriate results.
Conversely, the results of the fourth-order Trotter decomposition indicate that a bond dimension of $\chi=3000$ provides sufficient accuracy for analyzing the optical-conductivity spectra with finite frequency qualitatively.
It is worth noting that achieving full numerical convergence for the Drude weight is challenging since it necessitates long-time simulations maintaining high accuracy.

\section{TARPES and the Green's function}

By definition, Eq.~\eqref{eq:TrARPES} equals to
\begin{align}
  &A^{-} (k, \omega, \Delta t_{\mathrm{pr}})
  \notag \\
    &= \frac{1}{L} \sum_{j,\ell,\sigma} e^{-ik(r_j - r_{\ell})}
      \int^{\infty}_{-\infty} \dd{t_1}
      \int^{\infty}_{-\infty} \dd{t_2}
      e^{i\omega(t_1 - t_2)}
  \notag \\
    &\quad \times
      s(t_1 - t_{\mathrm{pr}}) s(t_2 - t_{\mathrm{pr}})
      \mel{\psi(t_2)}{\hat{c}^{\dagger}_{\ell,\sigma} \hat{U}(t_2,t_1) \hat{c}_{j,\sigma}}{\psi(t_1)}.
\end{align}
Upon partitioning the integration range of $t_2$ into two regimes, $t_2 > t_1$ and $t_2 < t_1$, we derive
\begin{widetext}
\begin{align}
  A^{-} (k, \omega, \Delta t_{\mathrm{pr}})
  &= \frac{1}{L} \sum_{j,\ell,\sigma} e^{-ik(r_j - r_{\ell})}
    \int^{\infty}_{-\infty} \dd{t_1}
    \int^{t_1}_{-\infty} \dd{t_2}
    e^{i\omega(t_1 - t_2)} s(t_1-t_{\mathrm{pr}}) s(t_2-t_{\mathrm{pr}})
    \mel{\psi(t_2)}{\hat{c}^{\dagger}_{\ell,\sigma} \hat{U}(t_2,t_1) \hat{c}_{j,\sigma}}{\psi(t_1)}
\notag \\
  &+ \frac{1}{L} \sum_{j,\ell,\sigma} e^{-ik(r_j - r_{\ell})}
  \int^{\infty}_{-\infty} \dd{t_1}
  \int^{\infty}_{t_1} \dd{t_2}
  e^{i\omega(t_1 - t_2)} s(t_1-t_{\mathrm{pr}}) s(t_2-t_{\mathrm{pr}})
  \mel{\psi(t_1)}{\hat{c}^{\dagger}_{j,\sigma} \hat{U}(t_1,t_2) \hat{c}_{\ell,\sigma}}{\psi(t_2)}^*,
\label{eq:qBbwTGJA}
\end{align}
and by replacing $\int^{\infty}_{-\infty} \dd{t_1} \int^{t_1}_{-\infty} \dd{t_2} = \int^{\infty}_{-\infty} \dd{t_2} \int^{\infty}_{t_2} \dd{t_1}$ in the first term, Eq.~\eqref{eq:qBbwTGJA} becomes
\begin{align}
  A^{-} (k, \omega, \Delta t_{\mathrm{pr}})
  &= \frac{1}{L} \sum_{j,\ell,\sigma} e^{-ik(r_j - r_{\ell})}
    \int^{\infty}_{-\infty} \dd{t_2}
    \int^{\infty}_{t_2} \dd{t_1}
    e^{i\omega(t_1 - t_2)} s(t_1-t_{\mathrm{pr}}) s(t_2-t_{\mathrm{pr}})
    \mel{\psi(t_2)}{\hat{c}^{\dagger}_{\ell,\sigma} \hat{U}(t_2,t_1) \hat{c}_{j,\sigma}}{\psi(t_1)}
\notag \\
  &+ \frac{1}{L} \sum_{j,\ell,\sigma} e^{-ik(r_j - r_{\ell})}
  \int^{\infty}_{-\infty} \dd{t_1}
  \int^{\infty}_{t_1} \dd{t_2}
  e^{i\omega(t_1 - t_2)} s(t_1-t_{\mathrm{pr}}) s(t_2-t_{\mathrm{pr}})
  \mel{\psi(t_1)}{\hat{c}^{\dagger}_{j,\sigma} \hat{U}(t_1,t_2) \hat{c}_{\ell,\sigma}}{\psi(t_2)}^*
\notag \\
  &= 2 \Im \qty[
    \frac{1}{L} \sum_{j,\ell,\sigma} e^{-ik(r_j - r_{\ell})}
    \int^{\infty}_{-\infty} \dd{t_2}
    \int^{\infty}_{t_2} \dd{t_1}
    e^{i\omega(t_1 - t_2)} s(t_1-t_{\mathrm{pr}}) s(t_2-t_{\mathrm{pr}})
    G^{<}_{j\ell}(t_1, t_2)
  ],
\label{eq:Jf27MuzE}
\end{align}
\end{widetext}
where $G^{<}_{j\ell}(t_1, t_2) = i \mel{\psi(t_2)}{\hat{c}^{\dagger}_{\ell,\sigma} \hat{U}(t_2,t_1) \hat{c}_{j,\sigma}}{\psi(t_1)}$ is the lesser Green's function.
Initially, we calculate a sequence of iMPS for each time by the iTEBD method.
Subsequently, we generate window states corresponding to $\hat{U}(t_1,t_2) \hat{c}_{\ell,\sigma}\ket{\psi(t_2)}$ and $\hat{c}_{j,\sigma} \ket{\psi(t_1)}$.
By evaluating the inner product of these states, we numerically obtain the integrand of Eq.~\eqref{eq:Jf27MuzE}.

\bibliography{paper.bbl}

\begin{thebibliography}{85}%
\makeatletter
\providecommand \@ifxundefined [1]{%
 \@ifx{#1\undefined}
}%
\providecommand \@ifnum [1]{%
 \ifnum #1\expandafter \@firstoftwo
 \else \expandafter \@secondoftwo
 \fi
}%
\providecommand \@ifx [1]{%
 \ifx #1\expandafter \@firstoftwo
 \else \expandafter \@secondoftwo
 \fi
}%
\providecommand \natexlab [1]{#1}%
\providecommand \enquote  [1]{``#1''}%
\providecommand \bibnamefont  [1]{#1}%
\providecommand \bibfnamefont [1]{#1}%
\providecommand \citenamefont [1]{#1}%
\providecommand \href@noop [0]{\@secondoftwo}%
\providecommand \href [0]{\begingroup \@sanitize@url \@href}%
\providecommand \@href[1]{\@@startlink{#1}\@@href}%
\providecommand \@@href[1]{\endgroup#1\@@endlink}%
\providecommand \@sanitize@url [0]{\catcode `\\12\catcode `\$12\catcode
  `\&12\catcode `\#12\catcode `\^12\catcode `\_12\catcode `\%12\relax}%
\providecommand \@@startlink[1]{}%
\providecommand \@@endlink[0]{}%
\providecommand \url  [0]{\begingroup\@sanitize@url \@url }%
\providecommand \@url [1]{\endgroup\@href {#1}{\urlprefix }}%
\providecommand \urlprefix  [0]{URL }%
\providecommand \Eprint [0]{\href }%
\providecommand \doibase [0]{http://dx.doi.org/}%
\providecommand \selectlanguage [0]{\@gobble}%
\providecommand \bibinfo  [0]{\@secondoftwo}%
\providecommand \bibfield  [0]{\@secondoftwo}%
\providecommand \translation [1]{[#1]}%
\providecommand \BibitemOpen [0]{}%
\providecommand \bibitemStop [0]{}%
\providecommand \bibitemNoStop [0]{.\EOS\space}%
\providecommand \EOS [0]{\spacefactor3000\relax}%
\providecommand \BibitemShut  [1]{\csname bibitem#1\endcsname}%
\let\auto@bib@innerbib\@empty
\bibitem [{\citenamefont {de~la Torre}\ \emph {et~al.}(2021)\citenamefont
  {de~la Torre}, \citenamefont {Kennes}, \citenamefont {Claassen},
  \citenamefont {Gerber}, \citenamefont {McIver},\ and\ \citenamefont
  {Sentef}}]{Torre2021RMP}%
  \BibitemOpen
  \bibfield  {author} {\bibinfo {author} {\bibfnamefont {A.}~\bibnamefont
  {de~la Torre}}, \bibinfo {author} {\bibfnamefont {D.~M.}\ \bibnamefont
  {Kennes}}, \bibinfo {author} {\bibfnamefont {M.}~\bibnamefont {Claassen}},
  \bibinfo {author} {\bibfnamefont {S.}~\bibnamefont {Gerber}}, \bibinfo
  {author} {\bibfnamefont {J.~W.}\ \bibnamefont {McIver}}, \ and\ \bibinfo
  {author} {\bibfnamefont {M.~A.}\ \bibnamefont {Sentef}},\ }\href {\doibase
  10.1103/RevModPhys.93.041002} {\bibfield  {journal} {\bibinfo  {journal}
  {Rev. Mod. Phys.}\ }\textbf {\bibinfo {volume} {93}},\ \bibinfo {pages}
  {041002} (\bibinfo {year} {2021})}\BibitemShut {NoStop}%
\bibitem [{\citenamefont {Dong}\ \emph {et~al.}(2022)\citenamefont {Dong},
  \citenamefont {Zhang},\ and\ \citenamefont {Wang}}]{Dong2022AM}%
  \BibitemOpen
  \bibfield  {author} {\bibinfo {author} {\bibfnamefont {T.}~\bibnamefont
  {Dong}}, \bibinfo {author} {\bibfnamefont {S.-J.}\ \bibnamefont {Zhang}}, \
  and\ \bibinfo {author} {\bibfnamefont {N.-L.}\ \bibnamefont {Wang}},\ }\href
  {\doibase 10.1002/adma.202110068} {\bibfield  {journal} {\bibinfo  {journal}
  {Adv. Mater.}\ }\textbf {\bibinfo {volume} {2022}},\ \bibinfo {pages}
  {2110068} (\bibinfo {year} {2022})}\BibitemShut {NoStop}%
\bibitem [{\citenamefont {Kaiser}(2017)}]{Kaiser2017PS}%
  \BibitemOpen
  \bibfield  {author} {\bibinfo {author} {\bibfnamefont {S.}~\bibnamefont
  {Kaiser}},\ }\href {\doibase 10.1088/1402-4896/aa8201} {\bibfield  {journal}
  {\bibinfo  {journal} {Phys. Scr.}\ }\textbf {\bibinfo {volume} {92}},\
  \bibinfo {pages} {103001} (\bibinfo {year} {2017})}\BibitemShut {NoStop}%
\bibitem [{\citenamefont {Cavalleri}(2018)}]{Cavalleri2018CP}%
  \BibitemOpen
  \bibfield  {author} {\bibinfo {author} {\bibfnamefont {A.}~\bibnamefont
  {Cavalleri}},\ }\href {\doibase 10.1080/00107514.2017.1406623} {\bibfield
  {journal} {\bibinfo  {journal} {Contemp. Phys.}\ }\textbf {\bibinfo {volume}
  {59}},\ \bibinfo {pages} {31} (\bibinfo {year} {2018})}\BibitemShut {NoStop}%
\bibitem [{\citenamefont {Fausti}\ \emph {et~al.}(2011)\citenamefont {Fausti},
  \citenamefont {Tobey}, \citenamefont {Dean}, \citenamefont {Kaiser},
  \citenamefont {Dienst}, \citenamefont {Hoffmann}, \citenamefont {Pyon},
  \citenamefont {Takayama}, \citenamefont {Takagi},\ and\ \citenamefont
  {Cavalleri}}]{Fausti2011S}%
  \BibitemOpen
  \bibfield  {author} {\bibinfo {author} {\bibfnamefont {D.}~\bibnamefont
  {Fausti}}, \bibinfo {author} {\bibfnamefont {R.~I.}\ \bibnamefont {Tobey}},
  \bibinfo {author} {\bibfnamefont {N.}~\bibnamefont {Dean}}, \bibinfo {author}
  {\bibfnamefont {S.}~\bibnamefont {Kaiser}}, \bibinfo {author} {\bibfnamefont
  {A.}~\bibnamefont {Dienst}}, \bibinfo {author} {\bibfnamefont {M.~C.}\
  \bibnamefont {Hoffmann}}, \bibinfo {author} {\bibfnamefont {S.}~\bibnamefont
  {Pyon}}, \bibinfo {author} {\bibfnamefont {T.}~\bibnamefont {Takayama}},
  \bibinfo {author} {\bibfnamefont {H.}~\bibnamefont {Takagi}}, \ and\ \bibinfo
  {author} {\bibfnamefont {A.}~\bibnamefont {Cavalleri}},\ }\href {\doibase
  10.1126/science.1197294} {\bibfield  {journal} {\bibinfo  {journal}
  {Science}\ }\textbf {\bibinfo {volume} {331}},\ \bibinfo {pages} {189}
  (\bibinfo {year} {2011})}\BibitemShut {NoStop}%
\bibitem [{\citenamefont {Suzuki}\ \emph {et~al.}(2019)\citenamefont {Suzuki},
  \citenamefont {Someya}, \citenamefont {Hashimoto}, \citenamefont {Michimae},
  \citenamefont {Watanabe}, \citenamefont {Fujisawa}, \citenamefont {Kanai},
  \citenamefont {Ishii}, \citenamefont {Itatani}, \citenamefont {Kasahara},
  \citenamefont {Matsuda}, \citenamefont {Shibauchi}, \citenamefont {Okazaki},\
  and\ \citenamefont {Shin}}]{Suzuki2019CP}%
  \BibitemOpen
  \bibfield  {author} {\bibinfo {author} {\bibfnamefont {T.}~\bibnamefont
  {Suzuki}}, \bibinfo {author} {\bibfnamefont {T.}~\bibnamefont {Someya}},
  \bibinfo {author} {\bibfnamefont {T.}~\bibnamefont {Hashimoto}}, \bibinfo
  {author} {\bibfnamefont {S.}~\bibnamefont {Michimae}}, \bibinfo {author}
  {\bibfnamefont {M.}~\bibnamefont {Watanabe}}, \bibinfo {author}
  {\bibfnamefont {M.}~\bibnamefont {Fujisawa}}, \bibinfo {author}
  {\bibfnamefont {T.}~\bibnamefont {Kanai}}, \bibinfo {author} {\bibfnamefont
  {N.}~\bibnamefont {Ishii}}, \bibinfo {author} {\bibfnamefont
  {J.}~\bibnamefont {Itatani}}, \bibinfo {author} {\bibfnamefont
  {S.}~\bibnamefont {Kasahara}}, \bibinfo {author} {\bibfnamefont
  {Y.}~\bibnamefont {Matsuda}}, \bibinfo {author} {\bibfnamefont
  {T.}~\bibnamefont {Shibauchi}}, \bibinfo {author} {\bibfnamefont
  {K.}~\bibnamefont {Okazaki}}, \ and\ \bibinfo {author} {\bibfnamefont
  {S.}~\bibnamefont {Shin}},\ }\href {\doibase 10.1038/s42005-019-0219-4}
  {\bibfield  {journal} {\bibinfo  {journal} {Commun. Phys.}\ }\textbf
  {\bibinfo {volume} {2}},\ \bibinfo {pages} {115} (\bibinfo {year}
  {2019})}\BibitemShut {NoStop}%
\bibitem [{\citenamefont {Isoyama}\ \emph {et~al.}(2021)\citenamefont
  {Isoyama}, \citenamefont {Yoshikawa}, \citenamefont {Katsumi}, \citenamefont
  {Wong}, \citenamefont {Shikama}, \citenamefont {Sakishita}, \citenamefont
  {Nabeshima}, \citenamefont {Maeda},\ and\ \citenamefont
  {Shimano}}]{Isoyama2021CP}%
  \BibitemOpen
  \bibfield  {author} {\bibinfo {author} {\bibfnamefont {K.}~\bibnamefont
  {Isoyama}}, \bibinfo {author} {\bibfnamefont {N.}~\bibnamefont {Yoshikawa}},
  \bibinfo {author} {\bibfnamefont {K.}~\bibnamefont {Katsumi}}, \bibinfo
  {author} {\bibfnamefont {J.}~\bibnamefont {Wong}}, \bibinfo {author}
  {\bibfnamefont {N.}~\bibnamefont {Shikama}}, \bibinfo {author} {\bibfnamefont
  {Y.}~\bibnamefont {Sakishita}}, \bibinfo {author} {\bibfnamefont
  {F.}~\bibnamefont {Nabeshima}}, \bibinfo {author} {\bibfnamefont
  {A.}~\bibnamefont {Maeda}}, \ and\ \bibinfo {author} {\bibfnamefont
  {R.}~\bibnamefont {Shimano}},\ }\href {\doibase 10.1038/s42005-021-00663-8}
  {\bibfield  {journal} {\bibinfo  {journal} {Commun. Phys.}\ }\textbf
  {\bibinfo {volume} {4}},\ \bibinfo {pages} {160} (\bibinfo {year}
  {2021})}\BibitemShut {NoStop}%
\bibitem [{\citenamefont {Buzzi}\ \emph {et~al.}(2020)\citenamefont {Buzzi},
  \citenamefont {Nicoletti}, \citenamefont {Fechner}, \citenamefont
  {Tancogne-Dejean}, \citenamefont {Sentef}, \citenamefont {Georges},
  \citenamefont {Biesner}, \citenamefont {Uykur}, \citenamefont {Dressel},
  \citenamefont {Henderson}, \citenamefont {Siegrist}, \citenamefont
  {Schlueter}, \citenamefont {Miyagawa}, \citenamefont {Kanoda}, \citenamefont
  {Nam}, \citenamefont {Ardavan}, \citenamefont {Coulthard}, \citenamefont
  {Tindall}, \citenamefont {Schlawin}, \citenamefont {Jaksch},\ and\
  \citenamefont {Cavalleri}}]{Buzzi2020PRX}%
  \BibitemOpen
  \bibfield  {author} {\bibinfo {author} {\bibfnamefont {M.}~\bibnamefont
  {Buzzi}}, \bibinfo {author} {\bibfnamefont {D.}~\bibnamefont {Nicoletti}},
  \bibinfo {author} {\bibfnamefont {M.}~\bibnamefont {Fechner}}, \bibinfo
  {author} {\bibfnamefont {N.}~\bibnamefont {Tancogne-Dejean}}, \bibinfo
  {author} {\bibfnamefont {M.~A.}\ \bibnamefont {Sentef}}, \bibinfo {author}
  {\bibfnamefont {A.}~\bibnamefont {Georges}}, \bibinfo {author} {\bibfnamefont
  {T.}~\bibnamefont {Biesner}}, \bibinfo {author} {\bibfnamefont
  {E.}~\bibnamefont {Uykur}}, \bibinfo {author} {\bibfnamefont
  {M.}~\bibnamefont {Dressel}}, \bibinfo {author} {\bibfnamefont
  {A.}~\bibnamefont {Henderson}}, \bibinfo {author} {\bibfnamefont
  {T.}~\bibnamefont {Siegrist}}, \bibinfo {author} {\bibfnamefont {J.~A.}\
  \bibnamefont {Schlueter}}, \bibinfo {author} {\bibfnamefont {K.}~\bibnamefont
  {Miyagawa}}, \bibinfo {author} {\bibfnamefont {K.}~\bibnamefont {Kanoda}},
  \bibinfo {author} {\bibfnamefont {M.-S.}\ \bibnamefont {Nam}}, \bibinfo
  {author} {\bibfnamefont {A.}~\bibnamefont {Ardavan}}, \bibinfo {author}
  {\bibfnamefont {J.}~\bibnamefont {Coulthard}}, \bibinfo {author}
  {\bibfnamefont {J.}~\bibnamefont {Tindall}}, \bibinfo {author} {\bibfnamefont
  {F.}~\bibnamefont {Schlawin}}, \bibinfo {author} {\bibfnamefont
  {D.}~\bibnamefont {Jaksch}}, \ and\ \bibinfo {author} {\bibfnamefont
  {A.}~\bibnamefont {Cavalleri}},\ }\href {\doibase 10.1103/PhysRevX.10.031028}
  {\bibfield  {journal} {\bibinfo  {journal} {Phys. Rev. X}\ }\textbf {\bibinfo
  {volume} {10}},\ \bibinfo {pages} {031028} (\bibinfo {year}
  {2020})}\BibitemShut {NoStop}%
\bibitem [{\citenamefont {Werner}\ \emph {et~al.}(2018)\citenamefont {Werner},
  \citenamefont {Strand}, \citenamefont {Hoshino}, \citenamefont {Murakami},\
  and\ \citenamefont {Eckstein}}]{Werner2018PRB}%
  \BibitemOpen
  \bibfield  {author} {\bibinfo {author} {\bibfnamefont {P.}~\bibnamefont
  {Werner}}, \bibinfo {author} {\bibfnamefont {H.~U.~R.}\ \bibnamefont
  {Strand}}, \bibinfo {author} {\bibfnamefont {S.}~\bibnamefont {Hoshino}},
  \bibinfo {author} {\bibfnamefont {Y.}~\bibnamefont {Murakami}}, \ and\
  \bibinfo {author} {\bibfnamefont {M.}~\bibnamefont {Eckstein}},\ }\href
  {\doibase 10.1103/PhysRevB.97.165119} {\bibfield  {journal} {\bibinfo
  {journal} {Phys. Rev. B}\ }\textbf {\bibinfo {volume} {97}},\ \bibinfo
  {pages} {165119} (\bibinfo {year} {2018})}\BibitemShut {NoStop}%
\bibitem [{\citenamefont {Wang}\ \emph {et~al.}(2018)\citenamefont {Wang},
  \citenamefont {Chen}, \citenamefont {Moritz},\ and\ \citenamefont
  {Devereaux}}]{Wang2018PRL}%
  \BibitemOpen
  \bibfield  {author} {\bibinfo {author} {\bibfnamefont {Y.}~\bibnamefont
  {Wang}}, \bibinfo {author} {\bibfnamefont {C.-C.}\ \bibnamefont {Chen}},
  \bibinfo {author} {\bibfnamefont {B.}~\bibnamefont {Moritz}}, \ and\ \bibinfo
  {author} {\bibfnamefont {T.~P.}\ \bibnamefont {Devereaux}},\ }\href {\doibase
  10.1103/PhysRevLett.120.246402} {\bibfield  {journal} {\bibinfo  {journal}
  {Phys. Rev. Lett.}\ }\textbf {\bibinfo {volume} {120}},\ \bibinfo {pages}
  {246402} (\bibinfo {year} {2018})}\BibitemShut {NoStop}%
\bibitem [{\citenamefont {Bittner}\ \emph {et~al.}(2019)\citenamefont
  {Bittner}, \citenamefont {Tohyama}, \citenamefont {Kaiser},\ and\
  \citenamefont {Manske}}]{Bittner2019JPSJ}%
  \BibitemOpen
  \bibfield  {author} {\bibinfo {author} {\bibfnamefont {N.}~\bibnamefont
  {Bittner}}, \bibinfo {author} {\bibfnamefont {T.}~\bibnamefont {Tohyama}},
  \bibinfo {author} {\bibfnamefont {S.}~\bibnamefont {Kaiser}}, \ and\ \bibinfo
  {author} {\bibfnamefont {D.}~\bibnamefont {Manske}},\ }\href {\doibase
  10.7566/JPSJ.88.044704} {\bibfield  {journal} {\bibinfo  {journal} {J. Phys.
  Soc. Jpn.}\ }\textbf {\bibinfo {volume} {88}},\ \bibinfo {pages} {044704}
  (\bibinfo {year} {2019})}\BibitemShut {NoStop}%
\bibitem [{\citenamefont {Schlawin}\ \emph {et~al.}(2019)\citenamefont
  {Schlawin}, \citenamefont {Cavalleri},\ and\ \citenamefont
  {Jaksch}}]{Schlawin2019PRL}%
  \BibitemOpen
  \bibfield  {author} {\bibinfo {author} {\bibfnamefont {F.}~\bibnamefont
  {Schlawin}}, \bibinfo {author} {\bibfnamefont {A.}~\bibnamefont {Cavalleri}},
  \ and\ \bibinfo {author} {\bibfnamefont {D.}~\bibnamefont {Jaksch}},\ }\href
  {\doibase 10.1103/PhysRevLett.122.133602} {\bibfield  {journal} {\bibinfo
  {journal} {Phys. Rev. Lett.}\ }\textbf {\bibinfo {volume} {122}},\ \bibinfo
  {pages} {133602} (\bibinfo {year} {2019})}\BibitemShut {NoStop}%
\bibitem [{\citenamefont {Werner}\ \emph {et~al.}(2019)\citenamefont {Werner},
  \citenamefont {Li}, \citenamefont {Gole\ifmmode~\check{z}\else \v{z}\fi{}},\
  and\ \citenamefont {Eckstein}}]{Werner2019PRB}%
  \BibitemOpen
  \bibfield  {author} {\bibinfo {author} {\bibfnamefont {P.}~\bibnamefont
  {Werner}}, \bibinfo {author} {\bibfnamefont {J.}~\bibnamefont {Li}}, \bibinfo
  {author} {\bibfnamefont {D.}~\bibnamefont {Gole\ifmmode~\check{z}\else
  \v{z}\fi{}}}, \ and\ \bibinfo {author} {\bibfnamefont {M.}~\bibnamefont
  {Eckstein}},\ }\href {\doibase 10.1103/PhysRevB.100.155130} {\bibfield
  {journal} {\bibinfo  {journal} {Phys. Rev. B}\ }\textbf {\bibinfo {volume}
  {100}},\ \bibinfo {pages} {155130} (\bibinfo {year} {2019})}\BibitemShut
  {NoStop}%
\bibitem [{\citenamefont {Kirilyuk}\ \emph {et~al.}(2010)\citenamefont
  {Kirilyuk}, \citenamefont {Kimel},\ and\ \citenamefont
  {Rasing}}]{Kirilyuk2010RMP}%
  \BibitemOpen
  \bibfield  {author} {\bibinfo {author} {\bibfnamefont {A.}~\bibnamefont
  {Kirilyuk}}, \bibinfo {author} {\bibfnamefont {A.~V.}\ \bibnamefont {Kimel}},
  \ and\ \bibinfo {author} {\bibfnamefont {T.}~\bibnamefont {Rasing}},\ }\href
  {\doibase 10.1103/RevModPhys.82.2731} {\bibfield  {journal} {\bibinfo
  {journal} {Rev. Mod. Phys.}\ }\textbf {\bibinfo {volume} {82}},\ \bibinfo
  {pages} {2731} (\bibinfo {year} {2010})}\BibitemShut {NoStop}%
\bibitem [{\citenamefont {Ishihara}(2019)}]{Ishihara2019JPSJ}%
  \BibitemOpen
  \bibfield  {author} {\bibinfo {author} {\bibfnamefont {S.}~\bibnamefont
  {Ishihara}},\ }\href {\doibase 10.7566/JPSJ.88.072001} {\bibfield  {journal}
  {\bibinfo  {journal} {J. Phys. Soc. Jpn.}\ }\textbf {\bibinfo {volume}
  {88}},\ \bibinfo {pages} {072001} (\bibinfo {year} {2019})}\BibitemShut
  {NoStop}%
\bibitem [{\citenamefont {Mikhaylovskiy}\ \emph {et~al.}(2015)\citenamefont
  {Mikhaylovskiy}, \citenamefont {Hendry}, \citenamefont {Secchi},
  \citenamefont {Mentink}, \citenamefont {Eckstein}, \citenamefont {Wu},
  \citenamefont {Pisarev}, \citenamefont {Kruglyak}, \citenamefont
  {Katsnelson}, \citenamefont {Rasing},\ and\ \citenamefont
  {Kimel}}]{Mikhaylovskiy2015NC}%
  \BibitemOpen
  \bibfield  {author} {\bibinfo {author} {\bibfnamefont {R.}~\bibnamefont
  {Mikhaylovskiy}}, \bibinfo {author} {\bibfnamefont {E.}~\bibnamefont
  {Hendry}}, \bibinfo {author} {\bibfnamefont {A.}~\bibnamefont {Secchi}},
  \bibinfo {author} {\bibfnamefont {J.}~\bibnamefont {Mentink}}, \bibinfo
  {author} {\bibfnamefont {M.}~\bibnamefont {Eckstein}}, \bibinfo {author}
  {\bibfnamefont {A.}~\bibnamefont {Wu}}, \bibinfo {author} {\bibfnamefont
  {R.}~\bibnamefont {Pisarev}}, \bibinfo {author} {\bibfnamefont
  {V.}~\bibnamefont {Kruglyak}}, \bibinfo {author} {\bibfnamefont
  {M.}~\bibnamefont {Katsnelson}}, \bibinfo {author} {\bibfnamefont
  {T.}~\bibnamefont {Rasing}}, \ and\ \bibinfo {author} {\bibfnamefont
  {A.}~\bibnamefont {Kimel}},\ }\href {\doibase 10.1038/ncomms9190} {\bibfield
  {journal} {\bibinfo  {journal} {Nat. Commun.}\ }\textbf {\bibinfo {volume}
  {6}},\ \bibinfo {pages} {8190} (\bibinfo {year} {2015})}\BibitemShut
  {NoStop}%
\bibitem [{\citenamefont {Baierl}\ \emph {et~al.}(2016)\citenamefont {Baierl},
  \citenamefont {Hohenleutner}, \citenamefont {Kampfrath}, \citenamefont
  {Zvezdin}, \citenamefont {Kimel}, \citenamefont {Huber},\ and\ \citenamefont
  {Mikhaylovskiy}}]{Baierl2016NP}%
  \BibitemOpen
  \bibfield  {author} {\bibinfo {author} {\bibfnamefont {S.}~\bibnamefont
  {Baierl}}, \bibinfo {author} {\bibfnamefont {M.}~\bibnamefont
  {Hohenleutner}}, \bibinfo {author} {\bibfnamefont {T.}~\bibnamefont
  {Kampfrath}}, \bibinfo {author} {\bibfnamefont {A.~K.}\ \bibnamefont
  {Zvezdin}}, \bibinfo {author} {\bibfnamefont {A.~V.}\ \bibnamefont {Kimel}},
  \bibinfo {author} {\bibfnamefont {R.}~\bibnamefont {Huber}}, \ and\ \bibinfo
  {author} {\bibfnamefont {R.~V.}\ \bibnamefont {Mikhaylovskiy}},\ }\href
  {\doibase 10.1038/nphoton.2016.181} {\bibfield  {journal} {\bibinfo
  {journal} {Nat. Photon.}\ }\textbf {\bibinfo {volume} {10}},\ \bibinfo
  {pages} {715} (\bibinfo {year} {2016})}\BibitemShut {NoStop}%
\bibitem [{\citenamefont {Afanasiev}\ \emph {et~al.}(2019)\citenamefont
  {Afanasiev}, \citenamefont {Gatilova}, \citenamefont {Groenendijk},
  \citenamefont {Ivanov}, \citenamefont {Gibert}, \citenamefont {Gariglio},
  \citenamefont {Mentink}, \citenamefont {Li}, \citenamefont {Dasari},
  \citenamefont {Eckstein}, \citenamefont {Rasing}, \citenamefont {Caviglia},\
  and\ \citenamefont {Kimel}}]{Afanasiev2019PRX}%
  \BibitemOpen
  \bibfield  {author} {\bibinfo {author} {\bibfnamefont {D.}~\bibnamefont
  {Afanasiev}}, \bibinfo {author} {\bibfnamefont {A.}~\bibnamefont {Gatilova}},
  \bibinfo {author} {\bibfnamefont {D.~J.}\ \bibnamefont {Groenendijk}},
  \bibinfo {author} {\bibfnamefont {B.~A.}\ \bibnamefont {Ivanov}}, \bibinfo
  {author} {\bibfnamefont {M.}~\bibnamefont {Gibert}}, \bibinfo {author}
  {\bibfnamefont {S.}~\bibnamefont {Gariglio}}, \bibinfo {author}
  {\bibfnamefont {J.}~\bibnamefont {Mentink}}, \bibinfo {author} {\bibfnamefont
  {J.}~\bibnamefont {Li}}, \bibinfo {author} {\bibfnamefont {N.}~\bibnamefont
  {Dasari}}, \bibinfo {author} {\bibfnamefont {M.}~\bibnamefont {Eckstein}},
  \bibinfo {author} {\bibfnamefont {T.}~\bibnamefont {Rasing}}, \bibinfo
  {author} {\bibfnamefont {A.~D.}\ \bibnamefont {Caviglia}}, \ and\ \bibinfo
  {author} {\bibfnamefont {A.~V.}\ \bibnamefont {Kimel}},\ }\href {\doibase
  10.1103/PhysRevX.9.021020} {\bibfield  {journal} {\bibinfo  {journal} {Phys.
  Rev. X}\ }\textbf {\bibinfo {volume} {9}},\ \bibinfo {pages} {021020}
  (\bibinfo {year} {2019})}\BibitemShut {NoStop}%
\bibitem [{\citenamefont {Schlauderer}\ \emph {et~al.}(2019)\citenamefont
  {Schlauderer}, \citenamefont {Lange}, \citenamefont {Baierl}, \citenamefont
  {Ebnet}, \citenamefont {Schmid}, \citenamefont {Valovcin}, \citenamefont
  {Zvezdin}, \citenamefont {Kimel}, \citenamefont {Mikhaylovskiy},\ and\
  \citenamefont {Huber}}]{Schlauderer2019N}%
  \BibitemOpen
  \bibfield  {author} {\bibinfo {author} {\bibfnamefont {S.}~\bibnamefont
  {Schlauderer}}, \bibinfo {author} {\bibfnamefont {C.}~\bibnamefont {Lange}},
  \bibinfo {author} {\bibfnamefont {S.}~\bibnamefont {Baierl}}, \bibinfo
  {author} {\bibfnamefont {T.}~\bibnamefont {Ebnet}}, \bibinfo {author}
  {\bibfnamefont {C.~P.}\ \bibnamefont {Schmid}}, \bibinfo {author}
  {\bibfnamefont {D.~C.}\ \bibnamefont {Valovcin}}, \bibinfo {author}
  {\bibfnamefont {A.~K.}\ \bibnamefont {Zvezdin}}, \bibinfo {author}
  {\bibfnamefont {A.~V.}\ \bibnamefont {Kimel}}, \bibinfo {author}
  {\bibfnamefont {R.~V.}\ \bibnamefont {Mikhaylovskiy}}, \ and\ \bibinfo
  {author} {\bibfnamefont {R.}~\bibnamefont {Huber}},\ }\href {\doibase
  10.1038/s41586-019-1174-7} {\bibfield  {journal} {\bibinfo  {journal}
  {Nature}\ }\textbf {\bibinfo {volume} {569}},\ \bibinfo {pages} {383}
  (\bibinfo {year} {2019})}\BibitemShut {NoStop}%
\bibitem [{\citenamefont {Mentink}\ \emph {et~al.}(2015)\citenamefont
  {Mentink}, \citenamefont {Balzer},\ and\ \citenamefont
  {Eckstein}}]{Mentink2015NC}%
  \BibitemOpen
  \bibfield  {author} {\bibinfo {author} {\bibfnamefont {J.~H.}\ \bibnamefont
  {Mentink}}, \bibinfo {author} {\bibfnamefont {K.}~\bibnamefont {Balzer}}, \
  and\ \bibinfo {author} {\bibfnamefont {M.}~\bibnamefont {Eckstein}},\ }\href
  {\doibase 10.1038/ncomms7708} {\bibfield  {journal} {\bibinfo  {journal}
  {Nat. Commun.}\ }\textbf {\bibinfo {volume} {6}},\ \bibinfo {pages} {6708}
  (\bibinfo {year} {2015})}\BibitemShut {NoStop}%
\bibitem [{\citenamefont {Li}\ \emph {et~al.}(2018)\citenamefont {Li},
  \citenamefont {Strand}, \citenamefont {Werner},\ and\ \citenamefont
  {Eckstein}}]{Li2018NC}%
  \BibitemOpen
  \bibfield  {author} {\bibinfo {author} {\bibfnamefont {J.}~\bibnamefont
  {Li}}, \bibinfo {author} {\bibfnamefont {H.~U.~R.}\ \bibnamefont {Strand}},
  \bibinfo {author} {\bibfnamefont {P.}~\bibnamefont {Werner}}, \ and\ \bibinfo
  {author} {\bibfnamefont {M.}~\bibnamefont {Eckstein}},\ }\href {\doibase
  10.1038/s41467-018-07051-x} {\bibfield  {journal} {\bibinfo  {journal} {Nat.
  Commun.}\ }\textbf {\bibinfo {volume} {9}},\ \bibinfo {pages} {4581}
  (\bibinfo {year} {2018})}\BibitemShut {NoStop}%
\bibitem [{\citenamefont {Okamoto}\ \emph {et~al.}(2007)\citenamefont
  {Okamoto}, \citenamefont {Matsuzaki}, \citenamefont {Wakabayashi},
  \citenamefont {Takahashi},\ and\ \citenamefont {Hasegawa}}]{Okamoto2007PRL}%
  \BibitemOpen
  \bibfield  {author} {\bibinfo {author} {\bibfnamefont {H.}~\bibnamefont
  {Okamoto}}, \bibinfo {author} {\bibfnamefont {H.}~\bibnamefont {Matsuzaki}},
  \bibinfo {author} {\bibfnamefont {T.}~\bibnamefont {Wakabayashi}}, \bibinfo
  {author} {\bibfnamefont {Y.}~\bibnamefont {Takahashi}}, \ and\ \bibinfo
  {author} {\bibfnamefont {T.}~\bibnamefont {Hasegawa}},\ }\href {\doibase
  10.1103/PhysRevLett.98.037401} {\bibfield  {journal} {\bibinfo  {journal}
  {Phys. Rev. Lett.}\ }\textbf {\bibinfo {volume} {98}},\ \bibinfo {pages}
  {037401} (\bibinfo {year} {2007})}\BibitemShut {NoStop}%
\bibitem [{\citenamefont {Uemura}\ \emph {et~al.}(2008)\citenamefont {Uemura},
  \citenamefont {Matsuzaki}, \citenamefont {Takahashi}, \citenamefont
  {Hasegawa},\ and\ \citenamefont {Okamoto}}]{Uemura2008JPSJ}%
  \BibitemOpen
  \bibfield  {author} {\bibinfo {author} {\bibfnamefont {H.}~\bibnamefont
  {Uemura}}, \bibinfo {author} {\bibfnamefont {H.}~\bibnamefont {Matsuzaki}},
  \bibinfo {author} {\bibfnamefont {Y.}~\bibnamefont {Takahashi}}, \bibinfo
  {author} {\bibfnamefont {T.}~\bibnamefont {Hasegawa}}, \ and\ \bibinfo
  {author} {\bibfnamefont {H.}~\bibnamefont {Okamoto}},\ }\href {\doibase
  10.1143/JPSJ.77.113714} {\bibfield  {journal} {\bibinfo  {journal} {J. Phys.
  Soc. Jpn.}\ }\textbf {\bibinfo {volume} {77}},\ \bibinfo {pages} {113714}
  (\bibinfo {year} {2008})}\BibitemShut {NoStop}%
\bibitem [{\citenamefont {Wall}\ \emph {et~al.}(2011)\citenamefont {Wall},
  \citenamefont {Brida}, \citenamefont {Clark}, \citenamefont {Ehrke},
  \citenamefont {Jaksch}, \citenamefont {Ardavan}, \citenamefont {Bonora},
  \citenamefont {Uemura}, \citenamefont {Takahashi}, \citenamefont {Hasegawa},
  \citenamefont {Okamoto}, \citenamefont {Cerullo},\ and\ \citenamefont
  {Cavalleri}}]{Wall2011NP}%
  \BibitemOpen
  \bibfield  {author} {\bibinfo {author} {\bibfnamefont {S.}~\bibnamefont
  {Wall}}, \bibinfo {author} {\bibfnamefont {D.}~\bibnamefont {Brida}},
  \bibinfo {author} {\bibfnamefont {S.~R.}\ \bibnamefont {Clark}}, \bibinfo
  {author} {\bibfnamefont {H.~P.}\ \bibnamefont {Ehrke}}, \bibinfo {author}
  {\bibfnamefont {D.}~\bibnamefont {Jaksch}}, \bibinfo {author} {\bibfnamefont
  {A.}~\bibnamefont {Ardavan}}, \bibinfo {author} {\bibfnamefont
  {S.}~\bibnamefont {Bonora}}, \bibinfo {author} {\bibfnamefont
  {H.}~\bibnamefont {Uemura}}, \bibinfo {author} {\bibfnamefont
  {Y.}~\bibnamefont {Takahashi}}, \bibinfo {author} {\bibfnamefont
  {T.}~\bibnamefont {Hasegawa}}, \bibinfo {author} {\bibfnamefont
  {H.}~\bibnamefont {Okamoto}}, \bibinfo {author} {\bibfnamefont
  {G.}~\bibnamefont {Cerullo}}, \ and\ \bibinfo {author} {\bibfnamefont
  {A.}~\bibnamefont {Cavalleri}},\ }\href {\doibase 10.1038/nphys1831}
  {\bibfield  {journal} {\bibinfo  {journal} {Nat. Phys.}\ }\textbf {\bibinfo
  {volume} {7}},\ \bibinfo {pages} {114} (\bibinfo {year} {2011})}\BibitemShut
  {NoStop}%
\bibitem [{\citenamefont {Miyamoto}\ \emph {et~al.}(2019)\citenamefont
  {Miyamoto}, \citenamefont {Kakizaki}, \citenamefont {Terashige},
  \citenamefont {Hata}, \citenamefont {Yamakawa}, \citenamefont {Morimoto},
  \citenamefont {Takamura}, \citenamefont {Yada}, \citenamefont {Takahashi},
  \citenamefont {Hasegawa}, \citenamefont {Matsuzaki}, \citenamefont
  {Tohyama},\ and\ \citenamefont {Okamoto}}]{Miyamoto2019CP}%
  \BibitemOpen
  \bibfield  {author} {\bibinfo {author} {\bibfnamefont {T.}~\bibnamefont
  {Miyamoto}}, \bibinfo {author} {\bibfnamefont {T.}~\bibnamefont {Kakizaki}},
  \bibinfo {author} {\bibfnamefont {T.}~\bibnamefont {Terashige}}, \bibinfo
  {author} {\bibfnamefont {D.}~\bibnamefont {Hata}}, \bibinfo {author}
  {\bibfnamefont {H.}~\bibnamefont {Yamakawa}}, \bibinfo {author}
  {\bibfnamefont {T.}~\bibnamefont {Morimoto}}, \bibinfo {author}
  {\bibfnamefont {N.}~\bibnamefont {Takamura}}, \bibinfo {author}
  {\bibfnamefont {H.}~\bibnamefont {Yada}}, \bibinfo {author} {\bibfnamefont
  {Y.}~\bibnamefont {Takahashi}}, \bibinfo {author} {\bibfnamefont
  {T.}~\bibnamefont {Hasegawa}}, \bibinfo {author} {\bibfnamefont
  {H.}~\bibnamefont {Matsuzaki}}, \bibinfo {author} {\bibfnamefont
  {T.}~\bibnamefont {Tohyama}}, \ and\ \bibinfo {author} {\bibfnamefont
  {H.}~\bibnamefont {Okamoto}},\ }\href {\doibase 10.1038/s42005-019-0223-8}
  {\bibfield  {journal} {\bibinfo  {journal} {Commun. Phys.}\ }\textbf
  {\bibinfo {volume} {2}},\ \bibinfo {pages} {131} (\bibinfo {year}
  {2019})}\BibitemShut {NoStop}%
\bibitem [{\citenamefont {Takamura}\ \emph {et~al.}(2023)\citenamefont
  {Takamura}, \citenamefont {Miyamoto}, \citenamefont {Liang}, \citenamefont
  {Asada}, \citenamefont {Terashige}, \citenamefont {Takahashi}, \citenamefont
  {Hasegawa},\ and\ \citenamefont {Okamoto}}]{Takamura2023PRB}%
  \BibitemOpen
  \bibfield  {author} {\bibinfo {author} {\bibfnamefont {N.}~\bibnamefont
  {Takamura}}, \bibinfo {author} {\bibfnamefont {T.}~\bibnamefont {Miyamoto}},
  \bibinfo {author} {\bibfnamefont {S.}~\bibnamefont {Liang}}, \bibinfo
  {author} {\bibfnamefont {K.}~\bibnamefont {Asada}}, \bibinfo {author}
  {\bibfnamefont {T.}~\bibnamefont {Terashige}}, \bibinfo {author}
  {\bibfnamefont {Y.}~\bibnamefont {Takahashi}}, \bibinfo {author}
  {\bibfnamefont {T.}~\bibnamefont {Hasegawa}}, \ and\ \bibinfo {author}
  {\bibfnamefont {H.}~\bibnamefont {Okamoto}},\ }\href {\doibase
  10.1103/PhysRevB.107.085147} {\bibfield  {journal} {\bibinfo  {journal}
  {Phys. Rev. B}\ }\textbf {\bibinfo {volume} {107}},\ \bibinfo {pages}
  {085147} (\bibinfo {year} {2023})}\BibitemShut {NoStop}%
\bibitem [{\citenamefont {Iwai}\ \emph {et~al.}(2003)\citenamefont {Iwai},
  \citenamefont {Ono}, \citenamefont {Maeda}, \citenamefont {Matsuzaki},
  \citenamefont {Kishida}, \citenamefont {Okamoto},\ and\ \citenamefont
  {Tokura}}]{Iwai2003PRL}%
  \BibitemOpen
  \bibfield  {author} {\bibinfo {author} {\bibfnamefont {S.}~\bibnamefont
  {Iwai}}, \bibinfo {author} {\bibfnamefont {M.}~\bibnamefont {Ono}}, \bibinfo
  {author} {\bibfnamefont {A.}~\bibnamefont {Maeda}}, \bibinfo {author}
  {\bibfnamefont {H.}~\bibnamefont {Matsuzaki}}, \bibinfo {author}
  {\bibfnamefont {H.}~\bibnamefont {Kishida}}, \bibinfo {author} {\bibfnamefont
  {H.}~\bibnamefont {Okamoto}}, \ and\ \bibinfo {author} {\bibfnamefont
  {Y.}~\bibnamefont {Tokura}},\ }\href {\doibase 10.1103/PhysRevLett.91.057401}
  {\bibfield  {journal} {\bibinfo  {journal} {Phys. Rev. Lett.}\ }\textbf
  {\bibinfo {volume} {91}},\ \bibinfo {pages} {057401} (\bibinfo {year}
  {2003})}\BibitemShut {NoStop}%
\bibitem [{\citenamefont {Matsuzaki}\ \emph {et~al.}(2006)\citenamefont
  {Matsuzaki}, \citenamefont {Yamashita},\ and\ \citenamefont
  {Okamoto}}]{Matsuzaki2006JPSJ}%
  \BibitemOpen
  \bibfield  {author} {\bibinfo {author} {\bibfnamefont {H.}~\bibnamefont
  {Matsuzaki}}, \bibinfo {author} {\bibfnamefont {M.}~\bibnamefont
  {Yamashita}}, \ and\ \bibinfo {author} {\bibfnamefont {H.}~\bibnamefont
  {Okamoto}},\ }\href {\doibase 10.1143/JPSJ.75.123701} {\bibfield  {journal}
  {\bibinfo  {journal} {J. Phys. Soc. Jpn.}\ }\textbf {\bibinfo {volume}
  {75}},\ \bibinfo {pages} {123701} (\bibinfo {year} {2006})}\BibitemShut
  {NoStop}%
\bibitem [{\citenamefont {Matsuzaki}\ \emph {et~al.}(2014)\citenamefont
  {Matsuzaki}, \citenamefont {Iwata}, \citenamefont {Miyamoto}, \citenamefont
  {Terashige}, \citenamefont {Iwano}, \citenamefont {Takaishi}, \citenamefont
  {Takamura}, \citenamefont {Kumagai}, \citenamefont {Yamashita}, \citenamefont
  {Takahashi}, \citenamefont {Wakabayashi},\ and\ \citenamefont
  {Okamoto}}]{Matsuzaki2014PRL}%
  \BibitemOpen
  \bibfield  {author} {\bibinfo {author} {\bibfnamefont {H.}~\bibnamefont
  {Matsuzaki}}, \bibinfo {author} {\bibfnamefont {M.}~\bibnamefont {Iwata}},
  \bibinfo {author} {\bibfnamefont {T.}~\bibnamefont {Miyamoto}}, \bibinfo
  {author} {\bibfnamefont {T.}~\bibnamefont {Terashige}}, \bibinfo {author}
  {\bibfnamefont {K.}~\bibnamefont {Iwano}}, \bibinfo {author} {\bibfnamefont
  {S.}~\bibnamefont {Takaishi}}, \bibinfo {author} {\bibfnamefont
  {M.}~\bibnamefont {Takamura}}, \bibinfo {author} {\bibfnamefont
  {S.}~\bibnamefont {Kumagai}}, \bibinfo {author} {\bibfnamefont
  {M.}~\bibnamefont {Yamashita}}, \bibinfo {author} {\bibfnamefont
  {R.}~\bibnamefont {Takahashi}}, \bibinfo {author} {\bibfnamefont
  {Y.}~\bibnamefont {Wakabayashi}}, \ and\ \bibinfo {author} {\bibfnamefont
  {H.}~\bibnamefont {Okamoto}},\ }\href {\doibase
  10.1103/PhysRevLett.113.096403} {\bibfield  {journal} {\bibinfo  {journal}
  {Phys. Rev. Lett.}\ }\textbf {\bibinfo {volume} {113}},\ \bibinfo {pages}
  {096403} (\bibinfo {year} {2014})}\BibitemShut {NoStop}%
\bibitem [{\citenamefont {Gomi}\ \emph {et~al.}(2005)\citenamefont {Gomi},
  \citenamefont {Takahashi}, \citenamefont {Ueda}, \citenamefont {Itoh},\ and\
  \citenamefont {Aihara}}]{Gomi2005PRB}%
  \BibitemOpen
  \bibfield  {author} {\bibinfo {author} {\bibfnamefont {H.}~\bibnamefont
  {Gomi}}, \bibinfo {author} {\bibfnamefont {A.}~\bibnamefont {Takahashi}},
  \bibinfo {author} {\bibfnamefont {T.}~\bibnamefont {Ueda}}, \bibinfo {author}
  {\bibfnamefont {H.}~\bibnamefont {Itoh}}, \ and\ \bibinfo {author}
  {\bibfnamefont {M.}~\bibnamefont {Aihara}},\ }\href {\doibase
  10.1103/PhysRevB.71.045129} {\bibfield  {journal} {\bibinfo  {journal} {Phys.
  Rev. B}\ }\textbf {\bibinfo {volume} {71}},\ \bibinfo {pages} {045129}
  (\bibinfo {year} {2005})}\BibitemShut {NoStop}%
\bibitem [{\citenamefont {Al-Hassanieh}\ \emph {et~al.}(2008)\citenamefont
  {Al-Hassanieh}, \citenamefont {Reboredo}, \citenamefont {Feiguin},
  \citenamefont {Gonz\'alez},\ and\ \citenamefont
  {Dagotto}}]{Al-Hassanieh2008PRL}%
  \BibitemOpen
  \bibfield  {author} {\bibinfo {author} {\bibfnamefont {K.~A.}\ \bibnamefont
  {Al-Hassanieh}}, \bibinfo {author} {\bibfnamefont {F.~A.}\ \bibnamefont
  {Reboredo}}, \bibinfo {author} {\bibfnamefont {A.~E.}\ \bibnamefont
  {Feiguin}}, \bibinfo {author} {\bibfnamefont {I.}~\bibnamefont {Gonz\'alez}},
  \ and\ \bibinfo {author} {\bibfnamefont {E.}~\bibnamefont {Dagotto}},\ }\href
  {\doibase 10.1103/PhysRevLett.100.166403} {\bibfield  {journal} {\bibinfo
  {journal} {Phys. Rev. Lett.}\ }\textbf {\bibinfo {volume} {100}},\ \bibinfo
  {pages} {166403} (\bibinfo {year} {2008})}\BibitemShut {NoStop}%
\bibitem [{\citenamefont {Takahashi}\ \emph {et~al.}(2008)\citenamefont
  {Takahashi}, \citenamefont {Itoh},\ and\ \citenamefont
  {Aihara}}]{Takahashi2008PRB}%
  \BibitemOpen
  \bibfield  {author} {\bibinfo {author} {\bibfnamefont {A.}~\bibnamefont
  {Takahashi}}, \bibinfo {author} {\bibfnamefont {H.}~\bibnamefont {Itoh}}, \
  and\ \bibinfo {author} {\bibfnamefont {M.}~\bibnamefont {Aihara}},\ }\href
  {\doibase 10.1103/PhysRevB.77.205105} {\bibfield  {journal} {\bibinfo
  {journal} {Phys. Rev. B}\ }\textbf {\bibinfo {volume} {77}},\ \bibinfo
  {pages} {205105} (\bibinfo {year} {2008})}\BibitemShut {NoStop}%
\bibitem [{\citenamefont {Lu}\ \emph {et~al.}(2012)\citenamefont {Lu},
  \citenamefont {Sota}, \citenamefont {Matsueda}, \citenamefont
  {Bon\ifmmode~\check{c}\else \v{c}\fi{}a},\ and\ \citenamefont
  {Tohyama}}]{Lu2012PRL}%
  \BibitemOpen
  \bibfield  {author} {\bibinfo {author} {\bibfnamefont {H.}~\bibnamefont
  {Lu}}, \bibinfo {author} {\bibfnamefont {S.}~\bibnamefont {Sota}}, \bibinfo
  {author} {\bibfnamefont {H.}~\bibnamefont {Matsueda}}, \bibinfo {author}
  {\bibfnamefont {J.}~\bibnamefont {Bon\ifmmode~\check{c}\else \v{c}\fi{}a}}, \
  and\ \bibinfo {author} {\bibfnamefont {T.}~\bibnamefont {Tohyama}},\ }\href
  {\doibase 10.1103/PhysRevLett.109.197401} {\bibfield  {journal} {\bibinfo
  {journal} {Phys. Rev. Lett.}\ }\textbf {\bibinfo {volume} {109}},\ \bibinfo
  {pages} {197401} (\bibinfo {year} {2012})}\BibitemShut {NoStop}%
\bibitem [{\citenamefont {Shao}\ \emph {et~al.}(2019)\citenamefont {Shao},
  \citenamefont {Lu}, \citenamefont {Luo},\ and\ \citenamefont
  {Mondaini}}]{Shao2019PRB}%
  \BibitemOpen
  \bibfield  {author} {\bibinfo {author} {\bibfnamefont {C.}~\bibnamefont
  {Shao}}, \bibinfo {author} {\bibfnamefont {H.}~\bibnamefont {Lu}}, \bibinfo
  {author} {\bibfnamefont {H.-G.}\ \bibnamefont {Luo}}, \ and\ \bibinfo
  {author} {\bibfnamefont {R.}~\bibnamefont {Mondaini}},\ }\href {\doibase
  10.1103/PhysRevB.100.041114} {\bibfield  {journal} {\bibinfo  {journal}
  {Phys. Rev. B}\ }\textbf {\bibinfo {volume} {100}},\ \bibinfo {pages}
  {041114(R)} (\bibinfo {year} {2019})}\BibitemShut {NoStop}%
\bibitem [{\citenamefont {Murakami}\ \emph {et~al.}(2022)\citenamefont
  {Murakami}, \citenamefont {Takayoshi}, \citenamefont {Kaneko}, \citenamefont
  {Sun}, \citenamefont {Gole\v{z}}, \citenamefont {Millis},\ and\ \citenamefont
  {Werner}}]{Murakami2022CP}%
  \BibitemOpen
  \bibfield  {author} {\bibinfo {author} {\bibfnamefont {Y.}~\bibnamefont
  {Murakami}}, \bibinfo {author} {\bibfnamefont {S.}~\bibnamefont {Takayoshi}},
  \bibinfo {author} {\bibfnamefont {T.}~\bibnamefont {Kaneko}}, \bibinfo
  {author} {\bibfnamefont {Z.}~\bibnamefont {Sun}}, \bibinfo {author}
  {\bibfnamefont {D.}~\bibnamefont {Gole\v{z}}}, \bibinfo {author}
  {\bibfnamefont {A.~J.}\ \bibnamefont {Millis}}, \ and\ \bibinfo {author}
  {\bibfnamefont {P.}~\bibnamefont {Werner}},\ }\href {\doibase
  10.1038/s42005-021-00799-7} {\bibfield  {journal} {\bibinfo  {journal}
  {Commun. Phys.}\ }\textbf {\bibinfo {volume} {5}},\ \bibinfo {pages} {23}
  (\bibinfo {year} {2022})}\BibitemShut {NoStop}%
\bibitem [{\citenamefont {Murakami}\ \emph {et~al.}(2023)\citenamefont
  {Murakami}, \citenamefont {Takayoshi}, \citenamefont {Kaneko}, \citenamefont
  {L\"auchli},\ and\ \citenamefont {Werner}}]{Murakami2023PRL}%
  \BibitemOpen
  \bibfield  {author} {\bibinfo {author} {\bibfnamefont {Y.}~\bibnamefont
  {Murakami}}, \bibinfo {author} {\bibfnamefont {S.}~\bibnamefont {Takayoshi}},
  \bibinfo {author} {\bibfnamefont {T.}~\bibnamefont {Kaneko}}, \bibinfo
  {author} {\bibfnamefont {A.~M.}\ \bibnamefont {L\"auchli}}, \ and\ \bibinfo
  {author} {\bibfnamefont {P.}~\bibnamefont {Werner}},\ }\href {\doibase
  10.1103/PhysRevLett.130.106501} {\bibfield  {journal} {\bibinfo  {journal}
  {Phys. Rev. Lett.}\ }\textbf {\bibinfo {volume} {130}},\ \bibinfo {pages}
  {106501} (\bibinfo {year} {2023})}\BibitemShut {NoStop}%
\bibitem [{\citenamefont {Lu}\ \emph {et~al.}(2015)\citenamefont {Lu},
  \citenamefont {Shao}, \citenamefont {Bon\ifmmode~\check{c}\else \v{c}\fi{}a},
  \citenamefont {Manske},\ and\ \citenamefont {Tohyama}}]{Lu2015PRB}%
  \BibitemOpen
  \bibfield  {author} {\bibinfo {author} {\bibfnamefont {H.}~\bibnamefont
  {Lu}}, \bibinfo {author} {\bibfnamefont {C.}~\bibnamefont {Shao}}, \bibinfo
  {author} {\bibfnamefont {J.}~\bibnamefont {Bon\ifmmode~\check{c}\else
  \v{c}\fi{}a}}, \bibinfo {author} {\bibfnamefont {D.}~\bibnamefont {Manske}},
  \ and\ \bibinfo {author} {\bibfnamefont {T.}~\bibnamefont {Tohyama}},\ }\href
  {\doibase 10.1103/PhysRevB.91.245117} {\bibfield  {journal} {\bibinfo
  {journal} {Phys. Rev. B}\ }\textbf {\bibinfo {volume} {91}},\ \bibinfo
  {pages} {245117} (\bibinfo {year} {2015})}\BibitemShut {NoStop}%
\bibitem [{\citenamefont {Rinc\'on}\ and\ \citenamefont
  {Feiguin}(2021)}]{Rincon2021PRB}%
  \BibitemOpen
  \bibfield  {author} {\bibinfo {author} {\bibfnamefont {J.}~\bibnamefont
  {Rinc\'on}}\ and\ \bibinfo {author} {\bibfnamefont {A.~E.}\ \bibnamefont
  {Feiguin}},\ }\href {\doibase 10.1103/PhysRevB.104.085122} {\bibfield
  {journal} {\bibinfo  {journal} {Phys. Rev. B}\ }\textbf {\bibinfo {volume}
  {104}},\ \bibinfo {pages} {085122} (\bibinfo {year} {2021})}\BibitemShut
  {NoStop}%
\bibitem [{\citenamefont {Ejima}\ and\ \citenamefont
  {Nishimoto}(2007)}]{Ejima2007PRL}%
  \BibitemOpen
  \bibfield  {author} {\bibinfo {author} {\bibfnamefont {S.}~\bibnamefont
  {Ejima}}\ and\ \bibinfo {author} {\bibfnamefont {S.}~\bibnamefont
  {Nishimoto}},\ }\href {\doibase 10.1103/PhysRevLett.99.216403} {\bibfield
  {journal} {\bibinfo  {journal} {Phys. Rev. Lett.}\ }\textbf {\bibinfo
  {volume} {99}},\ \bibinfo {pages} {216403} (\bibinfo {year}
  {2007})}\BibitemShut {NoStop}%
\bibitem [{\citenamefont {Vidal}(2007)}]{Vidal2007PRL}%
  \BibitemOpen
  \bibfield  {author} {\bibinfo {author} {\bibfnamefont {G.}~\bibnamefont
  {Vidal}},\ }\href {\doibase 10.1103/PhysRevLett.98.070201} {\bibfield
  {journal} {\bibinfo  {journal} {Phys. Rev. Lett.}\ }\textbf {\bibinfo
  {volume} {98}},\ \bibinfo {pages} {070201} (\bibinfo {year}
  {2007})}\BibitemShut {NoStop}%
\bibitem [{\citenamefont {Or\'us}\ and\ \citenamefont
  {Vidal}(2008)}]{Orus2008PRB}%
  \BibitemOpen
  \bibfield  {author} {\bibinfo {author} {\bibfnamefont {R.}~\bibnamefont
  {Or\'us}}\ and\ \bibinfo {author} {\bibfnamefont {G.}~\bibnamefont {Vidal}},\
  }\href {\doibase 10.1103/PhysRevB.78.155117} {\bibfield  {journal} {\bibinfo
  {journal} {Phys. Rev. B}\ }\textbf {\bibinfo {volume} {78}},\ \bibinfo
  {pages} {155117} (\bibinfo {year} {2008})}\BibitemShut {NoStop}%
\bibitem [{\citenamefont {Shao}\ \emph {et~al.}(2016)\citenamefont {Shao},
  \citenamefont {Tohyama}, \citenamefont {Luo},\ and\ \citenamefont
  {Lu}}]{Shao2016PRB}%
  \BibitemOpen
  \bibfield  {author} {\bibinfo {author} {\bibfnamefont {C.}~\bibnamefont
  {Shao}}, \bibinfo {author} {\bibfnamefont {T.}~\bibnamefont {Tohyama}},
  \bibinfo {author} {\bibfnamefont {H.-G.}\ \bibnamefont {Luo}}, \ and\
  \bibinfo {author} {\bibfnamefont {H.}~\bibnamefont {Lu}},\ }\href {\doibase
  10.1103/PhysRevB.93.195144} {\bibfield  {journal} {\bibinfo  {journal} {Phys.
  Rev. B}\ }\textbf {\bibinfo {volume} {93}},\ \bibinfo {pages} {195144}
  (\bibinfo {year} {2016})}\BibitemShut {NoStop}%
\bibitem [{\citenamefont {Shinjo}\ and\ \citenamefont
  {Tohyama}(2018)}]{Shinjo2018PRB}%
  \BibitemOpen
  \bibfield  {author} {\bibinfo {author} {\bibfnamefont {K.}~\bibnamefont
  {Shinjo}}\ and\ \bibinfo {author} {\bibfnamefont {T.}~\bibnamefont
  {Tohyama}},\ }\href {\doibase 10.1103/PhysRevB.98.165103} {\bibfield
  {journal} {\bibinfo  {journal} {Phys. Rev. B}\ }\textbf {\bibinfo {volume}
  {98}},\ \bibinfo {pages} {165103} (\bibinfo {year} {2018})}\BibitemShut
  {NoStop}%
\bibitem [{\citenamefont {Shinjo}\ \emph {et~al.}(2022)\citenamefont {Shinjo},
  \citenamefont {Sota},\ and\ \citenamefont {Tohyama}}]{Shinjo2022PRR}%
  \BibitemOpen
  \bibfield  {author} {\bibinfo {author} {\bibfnamefont {K.}~\bibnamefont
  {Shinjo}}, \bibinfo {author} {\bibfnamefont {S.}~\bibnamefont {Sota}}, \ and\
  \bibinfo {author} {\bibfnamefont {T.}~\bibnamefont {Tohyama}},\ }\href
  {\doibase 10.1103/PhysRevResearch.4.L032019} {\bibfield  {journal} {\bibinfo
  {journal} {Phys. Rev. Res.}\ }\textbf {\bibinfo {volume} {4}},\ \bibinfo
  {pages} {L032019} (\bibinfo {year} {2022})}\BibitemShut {NoStop}%
\bibitem [{\citenamefont {Eckstein}\ and\ \citenamefont
  {Kollar}(2008)}]{Eckstein2008PRB}%
  \BibitemOpen
  \bibfield  {author} {\bibinfo {author} {\bibfnamefont {M.}~\bibnamefont
  {Eckstein}}\ and\ \bibinfo {author} {\bibfnamefont {M.}~\bibnamefont
  {Kollar}},\ }\href {\doibase 10.1103/PhysRevB.78.205119} {\bibfield
  {journal} {\bibinfo  {journal} {Phys. Rev. B}\ }\textbf {\bibinfo {volume}
  {78}},\ \bibinfo {pages} {205119} (\bibinfo {year} {2008})}\BibitemShut
  {NoStop}%
\bibitem [{\citenamefont {Kennes}\ \emph {et~al.}(2017)\citenamefont {Kennes},
  \citenamefont {Wilner}, \citenamefont {Reichman},\ and\ \citenamefont
  {Millis}}]{Kennes2017PRB}%
  \BibitemOpen
  \bibfield  {author} {\bibinfo {author} {\bibfnamefont {D.~M.}\ \bibnamefont
  {Kennes}}, \bibinfo {author} {\bibfnamefont {E.~Y.}\ \bibnamefont {Wilner}},
  \bibinfo {author} {\bibfnamefont {D.~R.}\ \bibnamefont {Reichman}}, \ and\
  \bibinfo {author} {\bibfnamefont {A.~J.}\ \bibnamefont {Millis}},\ }\href
  {\doibase 10.1103/PhysRevB.96.054506} {\bibfield  {journal} {\bibinfo
  {journal} {Phys. Rev. B}\ }\textbf {\bibinfo {volume} {96}},\ \bibinfo
  {pages} {054506} (\bibinfo {year} {2017})}\BibitemShut {NoStop}%
\bibitem [{\citenamefont {Jeckelmann}(2002)}]{Jeckelmann2002PRB}%
  \BibitemOpen
  \bibfield  {author} {\bibinfo {author} {\bibfnamefont {E.}~\bibnamefont
  {Jeckelmann}},\ }\href {\doibase 10.1103/PhysRevB.66.045114} {\bibfield
  {journal} {\bibinfo  {journal} {Phys. Rev. B}\ }\textbf {\bibinfo {volume}
  {66}},\ \bibinfo {pages} {045114} (\bibinfo {year} {2002})}\BibitemShut
  {NoStop}%
\bibitem [{\citenamefont {Jeckelmann}(2003)}]{Jeckelmann2003PRB}%
  \BibitemOpen
  \bibfield  {author} {\bibinfo {author} {\bibfnamefont {E.}~\bibnamefont
  {Jeckelmann}},\ }\href {\doibase 10.1103/PhysRevB.67.075106} {\bibfield
  {journal} {\bibinfo  {journal} {Phys. Rev. B}\ }\textbf {\bibinfo {volume}
  {67}},\ \bibinfo {pages} {075106} (\bibinfo {year} {2003})}\BibitemShut
  {NoStop}%
\bibitem [{\citenamefont {Gallagher}\ and\ \citenamefont
  {Mazumdar}(1997)}]{Gallagher1997PRB}%
  \BibitemOpen
  \bibfield  {author} {\bibinfo {author} {\bibfnamefont {F.~B.}\ \bibnamefont
  {Gallagher}}\ and\ \bibinfo {author} {\bibfnamefont {S.}~\bibnamefont
  {Mazumdar}},\ }\href {\doibase 10.1103/PhysRevB.56.15025} {\bibfield
  {journal} {\bibinfo  {journal} {Phys. Rev. B}\ }\textbf {\bibinfo {volume}
  {56}},\ \bibinfo {pages} {15025} (\bibinfo {year} {1997})}\BibitemShut
  {NoStop}%
\bibitem [{\citenamefont {Ryzhii}\ \emph {et~al.}(2007)\citenamefont {Ryzhii},
  \citenamefont {Ryzhii},\ and\ \citenamefont {Otsuji}}]{Ryzhii2007JAP}%
  \BibitemOpen
  \bibfield  {author} {\bibinfo {author} {\bibfnamefont {V.}~\bibnamefont
  {Ryzhii}}, \bibinfo {author} {\bibfnamefont {M.}~\bibnamefont {Ryzhii}}, \
  and\ \bibinfo {author} {\bibfnamefont {T.}~\bibnamefont {Otsuji}},\ }\href
  {\doibase 10.1063/1.2717566} {\bibfield  {journal} {\bibinfo  {journal} {J.
  Appl. Phys.}\ }\textbf {\bibinfo {volume} {101}},\ \bibinfo {pages} {083114}
  (\bibinfo {year} {2007})}\BibitemShut {NoStop}%
\bibitem [{\citenamefont {Mizuno}\ \emph {et~al.}(2000)\citenamefont {Mizuno},
  \citenamefont {Tsutsui}, \citenamefont {Tohyama},\ and\ \citenamefont
  {Maekawa}}]{Mizuno2003PRB}%
  \BibitemOpen
  \bibfield  {author} {\bibinfo {author} {\bibfnamefont {Y.}~\bibnamefont
  {Mizuno}}, \bibinfo {author} {\bibfnamefont {K.}~\bibnamefont {Tsutsui}},
  \bibinfo {author} {\bibfnamefont {T.}~\bibnamefont {Tohyama}}, \ and\
  \bibinfo {author} {\bibfnamefont {S.}~\bibnamefont {Maekawa}},\ }\href
  {\doibase 10.1103/PhysRevB.62.R4769} {\bibfield  {journal} {\bibinfo
  {journal} {Phys. Rev. B}\ }\textbf {\bibinfo {volume} {62}},\ \bibinfo
  {pages} {R4769} (\bibinfo {year} {2000})}\BibitemShut {NoStop}%
\bibitem [{\citenamefont {Matsueda}\ \emph {et~al.}(2004)\citenamefont
  {Matsueda}, \citenamefont {Tohyama},\ and\ \citenamefont
  {Maekawa}}]{Matsueda2004PRB}%
  \BibitemOpen
  \bibfield  {author} {\bibinfo {author} {\bibfnamefont {H.}~\bibnamefont
  {Matsueda}}, \bibinfo {author} {\bibfnamefont {T.}~\bibnamefont {Tohyama}}, \
  and\ \bibinfo {author} {\bibfnamefont {S.}~\bibnamefont {Maekawa}},\ }\href
  {\doibase 10.1103/PhysRevB.70.033102} {\bibfield  {journal} {\bibinfo
  {journal} {Phys. Rev. B}\ }\textbf {\bibinfo {volume} {70}},\ \bibinfo
  {pages} {033102} (\bibinfo {year} {2004})}\BibitemShut {NoStop}%
\bibitem [{\citenamefont {Yamaguchi}\ \emph {et~al.}(2021)\citenamefont
  {Yamaguchi}, \citenamefont {Iwano}, \citenamefont {Miyamoto}, \citenamefont
  {Takamura}, \citenamefont {Kida}, \citenamefont {Takahashi}, \citenamefont
  {Hasegawa},\ and\ \citenamefont {Okamoto}}]{Yamaguchi2021PRB}%
  \BibitemOpen
  \bibfield  {author} {\bibinfo {author} {\bibfnamefont {T.}~\bibnamefont
  {Yamaguchi}}, \bibinfo {author} {\bibfnamefont {K.}~\bibnamefont {Iwano}},
  \bibinfo {author} {\bibfnamefont {T.}~\bibnamefont {Miyamoto}}, \bibinfo
  {author} {\bibfnamefont {N.}~\bibnamefont {Takamura}}, \bibinfo {author}
  {\bibfnamefont {N.}~\bibnamefont {Kida}}, \bibinfo {author} {\bibfnamefont
  {Y.}~\bibnamefont {Takahashi}}, \bibinfo {author} {\bibfnamefont
  {T.}~\bibnamefont {Hasegawa}}, \ and\ \bibinfo {author} {\bibfnamefont
  {H.}~\bibnamefont {Okamoto}},\ }\href {\doibase 10.1103/PhysRevB.103.045124}
  {\bibfield  {journal} {\bibinfo  {journal} {Phys. Rev. B}\ }\textbf {\bibinfo
  {volume} {103}},\ \bibinfo {pages} {045124} (\bibinfo {year}
  {2021})}\BibitemShut {NoStop}%
\bibitem [{\citenamefont {Udono}\ \emph {et~al.}(2022)\citenamefont {Udono},
  \citenamefont {Sugimoto}, \citenamefont {Kaneko},\ and\ \citenamefont
  {Ohta}}]{Udono2022PRB}%
  \BibitemOpen
  \bibfield  {author} {\bibinfo {author} {\bibfnamefont {M.}~\bibnamefont
  {Udono}}, \bibinfo {author} {\bibfnamefont {K.}~\bibnamefont {Sugimoto}},
  \bibinfo {author} {\bibfnamefont {T.}~\bibnamefont {Kaneko}}, \ and\ \bibinfo
  {author} {\bibfnamefont {Y.}~\bibnamefont {Ohta}},\ }\href {\doibase
  10.1103/PhysRevB.105.L241108} {\bibfield  {journal} {\bibinfo  {journal}
  {Phys. Rev. B}\ }\textbf {\bibinfo {volume} {105}},\ \bibinfo {pages}
  {L241108} (\bibinfo {year} {2022})}\BibitemShut {NoStop}%
\bibitem [{\citenamefont {Udono}\ \emph {et~al.}(2023)\citenamefont {Udono},
  \citenamefont {Kaneko},\ and\ \citenamefont {Sugimoto}}]{Udono2023PRB}%
  \BibitemOpen
  \bibfield  {author} {\bibinfo {author} {\bibfnamefont {M.}~\bibnamefont
  {Udono}}, \bibinfo {author} {\bibfnamefont {T.}~\bibnamefont {Kaneko}}, \
  and\ \bibinfo {author} {\bibfnamefont {K.}~\bibnamefont {Sugimoto}},\ }\href
  {\doibase 10.1103/PhysRevB.108.L081304} {\bibfield  {journal} {\bibinfo
  {journal} {Phys. Rev. B}\ }\textbf {\bibinfo {volume} {108}},\ \bibinfo
  {pages} {L081304} (\bibinfo {year} {2023})}\BibitemShut {NoStop}%
\bibitem [{\citenamefont {Freericks}\ \emph {et~al.}(2009)\citenamefont
  {Freericks}, \citenamefont {Krishnamurthy},\ and\ \citenamefont
  {Pruschke}}]{Freericks2009PRL}%
  \BibitemOpen
  \bibfield  {author} {\bibinfo {author} {\bibfnamefont {J.~K.}\ \bibnamefont
  {Freericks}}, \bibinfo {author} {\bibfnamefont {H.~R.}\ \bibnamefont
  {Krishnamurthy}}, \ and\ \bibinfo {author} {\bibfnamefont {T.}~\bibnamefont
  {Pruschke}},\ }\href {\doibase 10.1103/PhysRevLett.102.136401} {\bibfield
  {journal} {\bibinfo  {journal} {Phys. Rev. Lett.}\ }\textbf {\bibinfo
  {volume} {102}},\ \bibinfo {pages} {136401} (\bibinfo {year}
  {2009})}\BibitemShut {NoStop}%
\bibitem [{\citenamefont {Freericks}\ \emph {et~al.}(2015)\citenamefont
  {Freericks}, \citenamefont {Krishnamurthy}, \citenamefont {Sentef},\ and\
  \citenamefont {Devereaux}}]{Freericks2015PS}%
  \BibitemOpen
  \bibfield  {author} {\bibinfo {author} {\bibfnamefont {J.~K.}\ \bibnamefont
  {Freericks}}, \bibinfo {author} {\bibfnamefont {H.~R.}\ \bibnamefont
  {Krishnamurthy}}, \bibinfo {author} {\bibfnamefont {M.~A.}\ \bibnamefont
  {Sentef}}, \ and\ \bibinfo {author} {\bibfnamefont {T.~P.}\ \bibnamefont
  {Devereaux}},\ }\href {\doibase 10.1088/0031-8949/2015/T165/014012}
  {\bibfield  {journal} {\bibinfo  {journal} {Phys. Scr.}\ }\textbf {\bibinfo
  {volume} {2015}},\ \bibinfo {pages} {014012} (\bibinfo {year}
  {2015})}\BibitemShut {NoStop}%
\bibitem [{\citenamefont {Zauner}\ \emph {et~al.}(2015)\citenamefont {Zauner},
  \citenamefont {Ganahl}, \citenamefont {Evertz},\ and\ \citenamefont
  {Nishino}}]{Zauner2015JPCM}%
  \BibitemOpen
  \bibfield  {author} {\bibinfo {author} {\bibfnamefont {V.}~\bibnamefont
  {Zauner}}, \bibinfo {author} {\bibfnamefont {M.}~\bibnamefont {Ganahl}},
  \bibinfo {author} {\bibfnamefont {H.~G.}\ \bibnamefont {Evertz}}, \ and\
  \bibinfo {author} {\bibfnamefont {T.}~\bibnamefont {Nishino}},\ }\href
  {\doibase 10.1088/0953-8984/27/42/425602} {\bibfield  {journal} {\bibinfo
  {journal} {J. Phys. Condens. Matter}\ }\textbf {\bibinfo {volume} {27}},\
  \bibinfo {pages} {425602} (\bibinfo {year} {2015})}\BibitemShut {NoStop}%
\bibitem [{\citenamefont {Ejima}\ \emph {et~al.}(2022)\citenamefont {Ejima},
  \citenamefont {Lange},\ and\ \citenamefont {Fehske}}]{Ejima2022PRR}%
  \BibitemOpen
  \bibfield  {author} {\bibinfo {author} {\bibfnamefont {S.}~\bibnamefont
  {Ejima}}, \bibinfo {author} {\bibfnamefont {F.}~\bibnamefont {Lange}}, \ and\
  \bibinfo {author} {\bibfnamefont {H.}~\bibnamefont {Fehske}},\ }\href
  {\doibase 10.1103/PhysRevResearch.4.L012012} {\bibfield  {journal} {\bibinfo
  {journal} {Phys. Rev. Res.}\ }\textbf {\bibinfo {volume} {4}},\ \bibinfo
  {pages} {L012012} (\bibinfo {year} {2022})}\BibitemShut {NoStop}%
\bibitem [{\citenamefont {Lieb}\ and\ \citenamefont {Wu}(1968)}]{Lieb1968PRL}%
  \BibitemOpen
  \bibfield  {author} {\bibinfo {author} {\bibfnamefont {E.~H.}\ \bibnamefont
  {Lieb}}\ and\ \bibinfo {author} {\bibfnamefont {F.~Y.}\ \bibnamefont {Wu}},\
  }\href {\doibase 10.1103/PhysRevLett.20.1445} {\bibfield  {journal} {\bibinfo
   {journal} {Phys. Rev. Lett.}\ }\textbf {\bibinfo {volume} {20}},\ \bibinfo
  {pages} {1445} (\bibinfo {year} {1968})}\BibitemShut {NoStop}%
\bibitem [{\citenamefont {Essler}\ \emph {et~al.}(2005)\citenamefont {Essler},
  \citenamefont {Frahm}, \citenamefont {G^^c3^^b6hmann}, \citenamefont
  {Kl^^c3^^bcmper},\ and\ \citenamefont {Korepin}}]{Essler2005}%
  \BibitemOpen
  \bibfield  {author} {\bibinfo {author} {\bibfnamefont {F.~H.~L.}\
  \bibnamefont {Essler}}, \bibinfo {author} {\bibfnamefont {H.}~\bibnamefont
  {Frahm}}, \bibinfo {author} {\bibfnamefont {F.}~\bibnamefont
  {G^^c3^^b6hmann}}, \bibinfo {author} {\bibfnamefont {A.}~\bibnamefont
  {Kl^^c3^^bcmper}}, \ and\ \bibinfo {author} {\bibfnamefont {V.~E.}\
  \bibnamefont {Korepin}},\ }\href {\doibase 10.1017/CBO9780511534843} {\emph
  {\bibinfo {title} {The One-Dimensional Hubbard Model}}}\ (\bibinfo
  {publisher} {Cambridge University Press},\ \bibinfo {year}
  {2005})\BibitemShut {NoStop}%
\bibitem [{\citenamefont {Benthien}\ and\ \citenamefont
  {Jeckelmann}(2007)}]{Benthien2007PRB}%
  \BibitemOpen
  \bibfield  {author} {\bibinfo {author} {\bibfnamefont {H.}~\bibnamefont
  {Benthien}}\ and\ \bibinfo {author} {\bibfnamefont {E.}~\bibnamefont
  {Jeckelmann}},\ }\href {\doibase 10.1103/PhysRevB.75.205128} {\bibfield
  {journal} {\bibinfo  {journal} {Phys. Rev. B}\ }\textbf {\bibinfo {volume}
  {75}},\ \bibinfo {pages} {205128} (\bibinfo {year} {2007})}\BibitemShut
  {NoStop}%
\bibitem [{\citenamefont {Ejima}\ \emph {et~al.}(2021)\citenamefont {Ejima},
  \citenamefont {Lange},\ and\ \citenamefont {Fehske}}]{Ejima2021SPP}%
  \BibitemOpen
  \bibfield  {author} {\bibinfo {author} {\bibfnamefont {S.}~\bibnamefont
  {Ejima}}, \bibinfo {author} {\bibfnamefont {F.}~\bibnamefont {Lange}}, \ and\
  \bibinfo {author} {\bibfnamefont {H.}~\bibnamefont {Fehske}},\ }\href
  {\doibase 10.21468/SciPostPhys.10.3.077} {\bibfield  {journal} {\bibinfo
  {journal} {SciPost Phys.}\ }\textbf {\bibinfo {volume} {10}},\ \bibinfo
  {pages} {077} (\bibinfo {year} {2021})}\BibitemShut {NoStop}%
\bibitem [{\citenamefont {Murakami}\ \emph {et~al.}(2021)\citenamefont
  {Murakami}, \citenamefont {Takayoshi}, \citenamefont {Koga},\ and\
  \citenamefont {Werner}}]{Murakami2021PRB}%
  \BibitemOpen
  \bibfield  {author} {\bibinfo {author} {\bibfnamefont {Y.}~\bibnamefont
  {Murakami}}, \bibinfo {author} {\bibfnamefont {S.}~\bibnamefont {Takayoshi}},
  \bibinfo {author} {\bibfnamefont {A.}~\bibnamefont {Koga}}, \ and\ \bibinfo
  {author} {\bibfnamefont {P.}~\bibnamefont {Werner}},\ }\href {\doibase
  10.1103/PhysRevB.103.035110} {\bibfield  {journal} {\bibinfo  {journal}
  {Phys. Rev. B}\ }\textbf {\bibinfo {volume} {103}},\ \bibinfo {pages}
  {035110} (\bibinfo {year} {2021})}\BibitemShut {NoStop}%
\bibitem [{\citenamefont {Kaneko}\ \emph {et~al.}(2019)\citenamefont {Kaneko},
  \citenamefont {Shirakawa}, \citenamefont {Sorella},\ and\ \citenamefont
  {Yunoki}}]{Kaneko2019PRL}%
  \BibitemOpen
  \bibfield  {author} {\bibinfo {author} {\bibfnamefont {T.}~\bibnamefont
  {Kaneko}}, \bibinfo {author} {\bibfnamefont {T.}~\bibnamefont {Shirakawa}},
  \bibinfo {author} {\bibfnamefont {S.}~\bibnamefont {Sorella}}, \ and\
  \bibinfo {author} {\bibfnamefont {S.}~\bibnamefont {Yunoki}},\ }\href
  {\doibase 10.1103/PhysRevLett.122.077002} {\bibfield  {journal} {\bibinfo
  {journal} {Phys. Rev. Lett.}\ }\textbf {\bibinfo {volume} {122}},\ \bibinfo
  {pages} {077002} (\bibinfo {year} {2019})}\BibitemShut {NoStop}%
\bibitem [{\citenamefont {Kaneko}\ \emph {et~al.}(2020)\citenamefont {Kaneko},
  \citenamefont {Yunoki},\ and\ \citenamefont {Millis}}]{Kaneko2020PRR}%
  \BibitemOpen
  \bibfield  {author} {\bibinfo {author} {\bibfnamefont {T.}~\bibnamefont
  {Kaneko}}, \bibinfo {author} {\bibfnamefont {S.}~\bibnamefont {Yunoki}}, \
  and\ \bibinfo {author} {\bibfnamefont {A.~J.}\ \bibnamefont {Millis}},\
  }\href {\doibase 10.1103/PhysRevResearch.2.032027} {\bibfield  {journal}
  {\bibinfo  {journal} {Phys. Rev. Res.}\ }\textbf {\bibinfo {volume} {2}},\
  \bibinfo {pages} {032027(R)} (\bibinfo {year} {2020})}\BibitemShut {NoStop}%
\bibitem [{\citenamefont {Ejima}\ \emph
  {et~al.}(2020{\natexlab{a}})\citenamefont {Ejima}, \citenamefont {Kaneko},
  \citenamefont {Lange}, \citenamefont {Yunoki},\ and\ \citenamefont
  {Fehske}}]{Ejima2020PRR}%
  \BibitemOpen
  \bibfield  {author} {\bibinfo {author} {\bibfnamefont {S.}~\bibnamefont
  {Ejima}}, \bibinfo {author} {\bibfnamefont {T.}~\bibnamefont {Kaneko}},
  \bibinfo {author} {\bibfnamefont {F.}~\bibnamefont {Lange}}, \bibinfo
  {author} {\bibfnamefont {S.}~\bibnamefont {Yunoki}}, \ and\ \bibinfo {author}
  {\bibfnamefont {H.}~\bibnamefont {Fehske}},\ }\href {\doibase
  10.1103/PhysRevResearch.2.032008} {\bibfield  {journal} {\bibinfo  {journal}
  {Phys. Rev. Res.}\ }\textbf {\bibinfo {volume} {2}},\ \bibinfo {pages}
  {032008(R)} (\bibinfo {year} {2020}{\natexlab{a}})}\BibitemShut {NoStop}%
\bibitem [{\citenamefont {Ejima}\ \emph
  {et~al.}(2020{\natexlab{b}})\citenamefont {Ejima}, \citenamefont {Kaneko},
  \citenamefont {Lange}, \citenamefont {Yunoki},\ and\ \citenamefont
  {Fehske}}]{Ejima2020JPSCP}%
  \BibitemOpen
  \bibfield  {author} {\bibinfo {author} {\bibfnamefont {S.}~\bibnamefont
  {Ejima}}, \bibinfo {author} {\bibfnamefont {T.}~\bibnamefont {Kaneko}},
  \bibinfo {author} {\bibfnamefont {F.}~\bibnamefont {Lange}}, \bibinfo
  {author} {\bibfnamefont {S.}~\bibnamefont {Yunoki}}, \ and\ \bibinfo {author}
  {\bibfnamefont {H.}~\bibnamefont {Fehske}},\ }\href {\doibase
  10.7566/JPSCP.30.011184} {\bibfield  {journal} {\bibinfo  {journal} {JPS
  Conf. Proc.}\ }\textbf {\bibinfo {volume} {30}},\ \bibinfo {pages} {011184}
  (\bibinfo {year} {2020}{\natexlab{b}})}\BibitemShut {NoStop}%
\bibitem [{\citenamefont {Yang}(1989)}]{Yang1989PRL}%
  \BibitemOpen
  \bibfield  {author} {\bibinfo {author} {\bibfnamefont {C.~N.}\ \bibnamefont
  {Yang}},\ }\href {\doibase 10.1103/PhysRevLett.63.2144} {\bibfield  {journal}
  {\bibinfo  {journal} {Phys. Rev. Lett.}\ }\textbf {\bibinfo {volume} {63}},\
  \bibinfo {pages} {2144} (\bibinfo {year} {1989})}\BibitemShut {NoStop}%
\bibitem [{\citenamefont {Nocera}\ \emph {et~al.}(2018)\citenamefont {Nocera},
  \citenamefont {Essler},\ and\ \citenamefont {Feiguin}}]{Nocera2018PRB}%
  \BibitemOpen
  \bibfield  {author} {\bibinfo {author} {\bibfnamefont {A.}~\bibnamefont
  {Nocera}}, \bibinfo {author} {\bibfnamefont {F.~H.~L.}\ \bibnamefont
  {Essler}}, \ and\ \bibinfo {author} {\bibfnamefont {A.~E.}\ \bibnamefont
  {Feiguin}},\ }\href {\doibase 10.1103/PhysRevB.97.045146} {\bibfield
  {journal} {\bibinfo  {journal} {Phys. Rev. B}\ }\textbf {\bibinfo {volume}
  {97}},\ \bibinfo {pages} {045146} (\bibinfo {year} {2018})}\BibitemShut
  {NoStop}%
\bibitem [{\citenamefont {Nishida}\ \emph {et~al.}(2020)\citenamefont
  {Nishida}, \citenamefont {Fujiuchi}, \citenamefont {Sugimoto},\ and\
  \citenamefont {Ohta}}]{Nishida2020JPSJ}%
  \BibitemOpen
  \bibfield  {author} {\bibinfo {author} {\bibfnamefont {H.}~\bibnamefont
  {Nishida}}, \bibinfo {author} {\bibfnamefont {R.}~\bibnamefont {Fujiuchi}},
  \bibinfo {author} {\bibfnamefont {K.}~\bibnamefont {Sugimoto}}, \ and\
  \bibinfo {author} {\bibfnamefont {Y.}~\bibnamefont {Ohta}},\ }\href {\doibase
  10.7566/JPSJ.89.023702} {\bibfield  {journal} {\bibinfo  {journal} {J. Phys.
  Soc. Jpn.}\ }\textbf {\bibinfo {volume} {89}},\ \bibinfo {pages} {023702}
  (\bibinfo {year} {2020})}\BibitemShut {NoStop}%
\bibitem [{\citenamefont {Wang}\ \emph {et~al.}(2017)\citenamefont {Wang},
  \citenamefont {Claassen}, \citenamefont {Moritz},\ and\ \citenamefont
  {Devereaux}}]{Wang2017PRB}%
  \BibitemOpen
  \bibfield  {author} {\bibinfo {author} {\bibfnamefont {Y.}~\bibnamefont
  {Wang}}, \bibinfo {author} {\bibfnamefont {M.}~\bibnamefont {Claassen}},
  \bibinfo {author} {\bibfnamefont {B.}~\bibnamefont {Moritz}}, \ and\ \bibinfo
  {author} {\bibfnamefont {T.~P.}\ \bibnamefont {Devereaux}},\ }\href {\doibase
  10.1103/PhysRevB.96.235142} {\bibfield  {journal} {\bibinfo  {journal} {Phys.
  Rev. B}\ }\textbf {\bibinfo {volume} {96}},\ \bibinfo {pages} {235142}
  (\bibinfo {year} {2017})}\BibitemShut {NoStop}%
\bibitem [{\citenamefont {Zawadzki}\ and\ \citenamefont
  {Feiguin}(2019)}]{Zawadzki2019PRB}%
  \BibitemOpen
  \bibfield  {author} {\bibinfo {author} {\bibfnamefont {K.}~\bibnamefont
  {Zawadzki}}\ and\ \bibinfo {author} {\bibfnamefont {A.~E.}\ \bibnamefont
  {Feiguin}},\ }\href {\doibase 10.1103/PhysRevB.100.195124} {\bibfield
  {journal} {\bibinfo  {journal} {Phys. Rev. B}\ }\textbf {\bibinfo {volume}
  {100}},\ \bibinfo {pages} {195124} (\bibinfo {year} {2019})}\BibitemShut
  {NoStop}%
\bibitem [{\citenamefont {Shao}\ \emph {et~al.}(2020)\citenamefont {Shao},
  \citenamefont {Tohyama}, \citenamefont {Luo},\ and\ \citenamefont
  {Lu}}]{Shao2020PRB}%
  \BibitemOpen
  \bibfield  {author} {\bibinfo {author} {\bibfnamefont {C.}~\bibnamefont
  {Shao}}, \bibinfo {author} {\bibfnamefont {T.}~\bibnamefont {Tohyama}},
  \bibinfo {author} {\bibfnamefont {H.-G.}\ \bibnamefont {Luo}}, \ and\
  \bibinfo {author} {\bibfnamefont {H.}~\bibnamefont {Lu}},\ }\href {\doibase
  10.1103/PhysRevB.101.045128} {\bibfield  {journal} {\bibinfo  {journal}
  {Phys. Rev. B}\ }\textbf {\bibinfo {volume} {101}},\ \bibinfo {pages}
  {045128} (\bibinfo {year} {2020})}\BibitemShut {NoStop}%
\bibitem [{\citenamefont {Shao}\ \emph {et~al.}(2022)\citenamefont {Shao},
  \citenamefont {Lu}, \citenamefont {Zhang}, \citenamefont {Yu}, \citenamefont
  {Tohyama},\ and\ \citenamefont {Lu}}]{Shao2022PRL}%
  \BibitemOpen
  \bibfield  {author} {\bibinfo {author} {\bibfnamefont {C.}~\bibnamefont
  {Shao}}, \bibinfo {author} {\bibfnamefont {H.}~\bibnamefont {Lu}}, \bibinfo
  {author} {\bibfnamefont {X.}~\bibnamefont {Zhang}}, \bibinfo {author}
  {\bibfnamefont {C.}~\bibnamefont {Yu}}, \bibinfo {author} {\bibfnamefont
  {T.}~\bibnamefont {Tohyama}}, \ and\ \bibinfo {author} {\bibfnamefont
  {R.}~\bibnamefont {Lu}},\ }\href {\doibase 10.1103/PhysRevLett.128.047401}
  {\bibfield  {journal} {\bibinfo  {journal} {Phys. Rev. Lett.}\ }\textbf
  {\bibinfo {volume} {128}},\ \bibinfo {pages} {047401} (\bibinfo {year}
  {2022})}\BibitemShut {NoStop}%
\bibitem [{\citenamefont {Bohrdt}\ \emph {et~al.}(2018)\citenamefont {Bohrdt},
  \citenamefont {Greif}, \citenamefont {Demler}, \citenamefont {Knap},\ and\
  \citenamefont {Grusdt}}]{Bohrdt2018PRB}%
  \BibitemOpen
  \bibfield  {author} {\bibinfo {author} {\bibfnamefont {A.}~\bibnamefont
  {Bohrdt}}, \bibinfo {author} {\bibfnamefont {D.}~\bibnamefont {Greif}},
  \bibinfo {author} {\bibfnamefont {E.}~\bibnamefont {Demler}}, \bibinfo
  {author} {\bibfnamefont {M.}~\bibnamefont {Knap}}, \ and\ \bibinfo {author}
  {\bibfnamefont {F.}~\bibnamefont {Grusdt}},\ }\href {\doibase
  10.1103/PhysRevB.97.125117} {\bibfield  {journal} {\bibinfo  {journal} {Phys.
  Rev. B}\ }\textbf {\bibinfo {volume} {97}},\ \bibinfo {pages} {125117}
  (\bibinfo {year} {2018})}\BibitemShut {NoStop}%
\bibitem [{\citenamefont {Cao}\ \emph {et~al.}(2019)\citenamefont {Cao},
  \citenamefont {Mazzone}, \citenamefont {Meyers}, \citenamefont {Hill},
  \citenamefont {Liu}, \citenamefont {Wall},\ and\ \citenamefont
  {Dean}}]{Cao2019PTRSA}%
  \BibitemOpen
  \bibfield  {author} {\bibinfo {author} {\bibfnamefont {Y.}~\bibnamefont
  {Cao}}, \bibinfo {author} {\bibfnamefont {D.~G.}\ \bibnamefont {Mazzone}},
  \bibinfo {author} {\bibfnamefont {D.}~\bibnamefont {Meyers}}, \bibinfo
  {author} {\bibfnamefont {J.~P.}\ \bibnamefont {Hill}}, \bibinfo {author}
  {\bibfnamefont {X.}~\bibnamefont {Liu}}, \bibinfo {author} {\bibfnamefont
  {S.}~\bibnamefont {Wall}}, \ and\ \bibinfo {author} {\bibfnamefont
  {M.~P.~M.}\ \bibnamefont {Dean}},\ }\href {\doibase 10.1098/rsta.2017.0480}
  {\bibfield  {journal} {\bibinfo  {journal} {Philos. Trans. Royal Soc. A}\
  }\textbf {\bibinfo {volume} {377}},\ \bibinfo {pages} {20170480} (\bibinfo
  {year} {2019})}\BibitemShut {NoStop}%
\bibitem [{\citenamefont {Mitrano}\ and\ \citenamefont
  {Wang}(2020)}]{Mitrano2020CP}%
  \BibitemOpen
  \bibfield  {author} {\bibinfo {author} {\bibfnamefont {M.}~\bibnamefont
  {Mitrano}}\ and\ \bibinfo {author} {\bibfnamefont {Y.}~\bibnamefont {Wang}},\
  }\href {\doibase 10.1038/s42005-020-00447-6} {\bibfield  {journal} {\bibinfo
  {journal} {Commun. Phys.}\ }\textbf {\bibinfo {volume} {3}},\ \bibinfo
  {pages} {184} (\bibinfo {year} {2020})}\BibitemShut {NoStop}%
\bibitem [{\citenamefont {Gel'mukhanov}\ \emph {et~al.}(2021)\citenamefont
  {Gel'mukhanov}, \citenamefont {Odelius}, \citenamefont {Polyutov},
  \citenamefont {F\"ohlisch},\ and\ \citenamefont
  {Kimberg}}]{Gelmukhanov2021RMP}%
  \BibitemOpen
  \bibfield  {author} {\bibinfo {author} {\bibfnamefont {F.}~\bibnamefont
  {Gel'mukhanov}}, \bibinfo {author} {\bibfnamefont {M.}~\bibnamefont
  {Odelius}}, \bibinfo {author} {\bibfnamefont {S.~P.}\ \bibnamefont
  {Polyutov}}, \bibinfo {author} {\bibfnamefont {A.}~\bibnamefont
  {F\"ohlisch}}, \ and\ \bibinfo {author} {\bibfnamefont {V.}~\bibnamefont
  {Kimberg}},\ }\href {\doibase 10.1103/RevModPhys.93.035001} {\bibfield
  {journal} {\bibinfo  {journal} {Rev. Mod. Phys.}\ }\textbf {\bibinfo {volume}
  {93}},\ \bibinfo {pages} {035001} (\bibinfo {year} {2021})}\BibitemShut
  {NoStop}%
\bibitem [{dlr()}]{dlr}%
  \BibitemOpen
  \href@noop {} {}\bibinfo {howpublished}
  {\url{https://qci.dlr.de/alqu/}}\BibitemShut {NoStop}%
\bibitem [{\citenamefont {Fishman}\ \emph {et~al.}(2022)\citenamefont
  {Fishman}, \citenamefont {White},\ and\ \citenamefont
  {Stoudenmire}}]{Fishman2022SPC}%
  \BibitemOpen
  \bibfield  {author} {\bibinfo {author} {\bibfnamefont {M.}~\bibnamefont
  {Fishman}}, \bibinfo {author} {\bibfnamefont {S.~R.}\ \bibnamefont {White}},
  \ and\ \bibinfo {author} {\bibfnamefont {E.~M.}\ \bibnamefont
  {Stoudenmire}},\ }\href {\doibase 10.21468/SciPostPhysCodeb.4} {\bibfield
  {journal} {\bibinfo  {journal} {SciPost Phys. Codebases}\ ,\ \bibinfo {pages}
  {4}} (\bibinfo {year} {2022})}\BibitemShut {NoStop}%
\bibitem [{\citenamefont {Segawa}\ \emph {et~al.}(2011)\citenamefont {Segawa},
  \citenamefont {Takahashi}, \citenamefont {Gomi},\ and\ \citenamefont
  {Aihara}}]{Segawa2011JPSJ}%
  \BibitemOpen
  \bibfield  {author} {\bibinfo {author} {\bibfnamefont {M.}~\bibnamefont
  {Segawa}}, \bibinfo {author} {\bibfnamefont {A.}~\bibnamefont {Takahashi}},
  \bibinfo {author} {\bibfnamefont {H.}~\bibnamefont {Gomi}}, \ and\ \bibinfo
  {author} {\bibfnamefont {M.}~\bibnamefont {Aihara}},\ }\href {\doibase
  10.1143/JPSJ.80.084721} {\bibfield  {journal} {\bibinfo  {journal} {J. Phys.
  Soc. Jpn.}\ }\textbf {\bibinfo {volume} {80}},\ \bibinfo {pages} {084721}
  (\bibinfo {year} {2011})}\BibitemShut {NoStop}%
\bibitem [{\citenamefont {Gomi}\ \emph {et~al.}(2014)\citenamefont {Gomi},
  \citenamefont {Hatano}, \citenamefont {Inagaki},\ and\ \citenamefont
  {Takahashi}}]{Gomi2014JPSJ}%
  \BibitemOpen
  \bibfield  {author} {\bibinfo {author} {\bibfnamefont {H.}~\bibnamefont
  {Gomi}}, \bibinfo {author} {\bibfnamefont {H.}~\bibnamefont {Hatano}},
  \bibinfo {author} {\bibfnamefont {T.~J.}\ \bibnamefont {Inagaki}}, \ and\
  \bibinfo {author} {\bibfnamefont {A.}~\bibnamefont {Takahashi}},\ }\href
  {\doibase 10.7566/JPSJ.83.094718} {\bibfield  {journal} {\bibinfo  {journal}
  {J. Phys. Soc. Jpn.}\ }\textbf {\bibinfo {volume} {83}},\ \bibinfo {pages}
  {094718} (\bibinfo {year} {2014})}\BibitemShut {NoStop}%
\bibitem [{\citenamefont {Lenar\ifmmode \check{c}\else
  \v{c}\fi{}i\ifmmode~\check{c}\else \v{c}\fi{}}\ \emph
  {et~al.}(2014)\citenamefont {Lenar\ifmmode \check{c}\else
  \v{c}\fi{}i\ifmmode~\check{c}\else \v{c}\fi{}}, \citenamefont
  {Gole\ifmmode~\check{z}\else \v{z}\fi{}}, \citenamefont
  {Bon\ifmmode~\check{c}\else \v{c}\fi{}a},\ and\ \citenamefont
  {Prelov\ifmmode~\check{s}\else \v{s}\fi{}ek}}]{Lenarcic2014PRB}%
  \BibitemOpen
  \bibfield  {author} {\bibinfo {author} {\bibfnamefont {Z.}~\bibnamefont
  {Lenar\ifmmode \check{c}\else \v{c}\fi{}i\ifmmode~\check{c}\else
  \v{c}\fi{}}}, \bibinfo {author} {\bibfnamefont {D.}~\bibnamefont
  {Gole\ifmmode~\check{z}\else \v{z}\fi{}}}, \bibinfo {author} {\bibfnamefont
  {J.}~\bibnamefont {Bon\ifmmode~\check{c}\else \v{c}\fi{}a}}, \ and\ \bibinfo
  {author} {\bibfnamefont {P.}~\bibnamefont {Prelov\ifmmode~\check{s}\else
  \v{s}\fi{}ek}},\ }\href {\doibase 10.1103/PhysRevB.89.125123} {\bibfield
  {journal} {\bibinfo  {journal} {Phys. Rev. B}\ }\textbf {\bibinfo {volume}
  {89}},\ \bibinfo {pages} {125123} (\bibinfo {year} {2014})}\BibitemShut
  {NoStop}%
\bibitem [{\citenamefont {Ohara}\ \emph {et~al.}(2013)\citenamefont {Ohara},
  \citenamefont {Kanamori},\ and\ \citenamefont {Ishihara}}]{Ohara2013PRB}%
  \BibitemOpen
  \bibfield  {author} {\bibinfo {author} {\bibfnamefont {J.}~\bibnamefont
  {Ohara}}, \bibinfo {author} {\bibfnamefont {Y.}~\bibnamefont {Kanamori}}, \
  and\ \bibinfo {author} {\bibfnamefont {S.}~\bibnamefont {Ishihara}},\ }\href
  {\doibase 10.1103/PhysRevB.88.085107} {\bibfield  {journal} {\bibinfo
  {journal} {Phys. Rev. B}\ }\textbf {\bibinfo {volume} {88}},\ \bibinfo
  {pages} {085107} (\bibinfo {year} {2013})}\BibitemShut {NoStop}%
\end{thebibliography}%

\end{document}